\documentclass{siamart0516}
\usepackage{euscript,amsmath,amssymb,amsfonts,graphicx,bm,mathtools}
  \usepackage{hyperref}
\usepackage{latexsym,amssymb,enumerate,amsmath} 
\usepackage{amsfonts,amsmath,latexsym,verbatim,amscd,mathrsfs,color,array,amssymb,graphicx}
\usepackage{subcaption}

  \usepackage{graphicx} 
  \usepackage{tikz}
\usepackage{fancyvrb}

\numberwithin{equation}{section}

\newcommand{\ep}{\varepsilon}
\newcommand{\Si}{\mathrm{Si}}
\newcommand{\hot}{\mathrm{h.o.t.}}
\newcommand{\diag}{\mathrm{diag}}
\newcommand{\pvint}{\mathrm{P.V.}\int}

\newtheorem{principal_result}{Principal Result}

\newtheorem{remark}{Remark}[section]

\newcommand{\dd}{\textup{d}}
\def\eps{\varepsilon}
\def\E{\mathbb{E}}

\def\R{\mathbb{R}}

\def\<{\langle}
\def\>{\rangle}
\usepackage{amsfonts}
\usepackage{graphicx}
\usepackage{epstopdf}
\usepackage{algorithmic}
\ifpdf
  \DeclareGraphicsExtensions{.eps,.pdf,.png,.jpg}
\else
  \DeclareGraphicsExtensions{.eps}
\fi

\newcommand{\TheTitle}{First Hitting Time of a One-Dimensional L\'evy Flight to Small Targets}
\newcommand{\ShortTitle}{L\'evy Flight First Hitting Time}
\newcommand{\TheAuthors}{Daniel Gomez and Sean D. Lawley}

\headers{\ShortTitle}{\TheAuthors}

\title{\TheTitle\thanks{
\funding{DG was supported by the Simons Foundation Math + X grant. SDL was supported by the National Science Foundation (DMS-1944574 and DMS-1814832).}}}

\author{Daniel Gomez\thanks{Center for Mathematical Biology \& Department of Mathematics,
University of Pennsylvania, Philadelphia, PA 19104, USA (\texttt{d1gomez@sas.upenn.edu}).} \and Sean D. Lawley\thanks{Department of Mathematics, University of Utah, Salt Lake City, UT 84112 USA (\texttt{lawley@math.utah.edu}).}% SDL was supported by the National Science Foundation (Grant Nos.\ CAREER DMS-1944574 and DMS-1814832).}
}
\date{\today}

\ifpdf
\hypersetup{
  pdftitle={\TheTitle},
  pdfauthor={\TheAuthors}
}
\fi

\begin{document}

\maketitle

\begin{abstract}
First hitting times (FHTs) describe the time it takes a random ``searcher'' to find a ``target'' and are used to study timescales in many applications. FHTs have been well-studied for diffusive search, especially for small targets, which is called the narrow capture or narrow escape problem. In this paper, we study the first hitting time to small targets for a one-dimensional superdiffusive search described by a L{\'e}vy flight. By applying the method of matched asymptotic expansions to a fractional differential equation we obtain an explicit asymptotic expansion for the mean FHT (MFHT). For fractional order $s\in(0,1)$ (describing a $(2s)$-stable L{\'e}vy flight whose squared displacement scales as $t^{1/s}$ in time $t$) and targets of radius $\eps\ll1$, we show that the MFHT is order one for $s\in(1/2,1)$ and diverges as $\log(1/\eps)$ for $s=1/2$ and $\eps^{2s-1}$ for $s\in(0,1/2)$. We then use our asymptotic results to identify the value of $s\in(0,1]$ which minimizes the average MFHT and find that (a) this optimal value of $s$ vanishes for sparse targets and (b) the value $s=1/2$ (corresponding to an inverse square L\'evy search) is optimal in only very specific circumstances. We confirm our results by comparison to both deterministic numerical solutions of the associated fractional differential equation and stochastic simulations.
\end{abstract}
%These three cases mirror the behavior of mean FHTs of diffusive (Brownian) search in respective dimensions one, two, and three, and similarly stem from differences in recurrence versus transience. 

% 60J60  	Diffusion processes 
% 35R11  	Fractional partial differential equations 
% 92C40  	Biochemistry, molecular biology

%\tableofcontents

%%%%%%%%%%%%%%%%%%%%%%%%%%%%%%%%%%%%%%%%%%%%%%%%%%%%%%%%%%%%%%%%%%%%%%%%%%%%%%%%%%%%%%%%%%%%%%%%%%%%%%%%%%%%%%%%%%%%%%%%%%%%%%%%%%%%%%%%%%%%%%%%%%%%%%%%%%%%%%%%%%%%%%%%%%%%%%%%%%%%%%%%%%%%%%%%%%%%%%%%%%%%%%%%%%%%%%%%%
\section{Introduction}\label{sec:intro}

The timescales of many physical, chemical, and biological processes are characterized by first hitting times (FHTs) \cite{benichou2011rev, redner2001, chou_first_2014, ralf2014}. Generically, a FHT is the time it takes a ``searcher'' to find a ``target.'' Applications include animal foraging \cite{viswanathan2008, kurella2015}, transcription factor search for DNA binding sites \cite{lomholt2005, mirny2009}, synaptic transmission in neuroscience \cite{reva2021}, menopause timing \cite{lawley2023bor}, financial income dynamics \cite{jolakoski2022}, and computer search algorithms \cite{pavlyukevich2007,pavlyukevich2008} among many other applications \cite{benichou2011rev, redner2001, chou_first_2014}. 
FHTs are often called first passage times, first arrival times, exit times, escape times, or capture times.

Mathematical models of such processes often assume that the searcher randomly explores a given spatial domain, and a great deal of mathematical and computational methods have been developed to study the statistics and probability distribution of the FHT to the target(s) \cite{benichou2010, holcman2014, vaccario2015, lindsay2017, bressloff2017mean, kaye2020}. More precisely, let $X=\{X(t)\}_{t\ge0}$ denote the stochastic path of a searcher in a $d$-dimensional spatial domain $\Omega\subseteq\R^{d}$. The FHT to a target set $\Omega_{\text{target}}\subset \Omega$ (where $\Omega_{\text{target}}$ is possibly a union of multiple disjoint sets) is then
\begin{align}\label{tau}
\tau
:=\inf\{t\geq0:X(t)\in \Omega_{\text{target}}\}.
\end{align}
Naturally, the statistics and distribution of the FHT $\tau$ depend on the stochastic dynamics of the searcher $X$, the space dimension $d\ge1$, and the size and geometry of the target set $\Omega_{\text{target}}$ and spatial domain $\Omega$. 

A common framework for studying FHTs is to assume that the searcher $X$ is a pure diffusion process (i.e.\ a Brownian motion) and the targets are much smaller than their confining spatial domain, which is called the narrow capture problem (or narrow escape problem if the target is embedded in the otherwise reflecting boundary) \cite{ward10, ward10b, Ammari2011, holcman2014, grebenkov2017narrow, bressloff2022}. For bounded domains in dimension $d=1$, the MFHT of such a diffusive searcher is always finite even if the targets are single points. In contrast, the MFHT of diffusion in any dimension $d\ge2$ diverges as the target size vanishes. In particular, if $\eps>0$ compares the lengthscale of the target to the lengthscale of the confining domain, then it is well-known that as $\eps$ vanishes,
\begin{align}\label{diff}
\E[\tau]
=\begin{cases}
O(1) &\text{if }d=1,\\
O(\log (1/\eps)) &\text{if }d=2,\\
O(\eps^{2-d}) &\text{if }d\ge3.
\end{cases}
\end{align}
The stark contrast in \eqref{diff} between dimensions $d=1$, $d=2$, and $d\ge3$ stems from the fact that Brownian motion is recurrent if $d=1$, neighborhood recurrent in $d=2$, and transient in $d\ge3$ \cite{durrett2019}.

%The FHT $\tau$ has been studied under a variety of assumptions on the stochastic dynamics of $X$ \cite{ }, with perhaps the most common assumption being that $X$ is a pure diffusion process (i.e.\ a Brownian motion). In this diffusion case, calculating the mean FHT is an elementary exercise in dimension $d=1$. In higher space dimensions, many prior works study the mean FHT for small targets, which has been dubbed the narrow capture problem or narrow escape problem \cite{lotsofnarrowstuffwardholcmanbenichougrebenkovbressloff}. 

FHTs have also been studied for superdiffusive processes, which are characterized by squared displacements that grow superlinearly in time \cite{metzler2004, chechkin2003, lomholt2008, palyulin2014, kusmierz2015, palyulin2016, padash2019, padash2020, wieschen2020, guinard2021, chaubet2022, padash2022, tzou2023}. A common mathematical model of superdiffusion is a L{\'e}vy flight \cite{bertoin1996, dubkov2008}, which is often derived from the continuous time random walk model \cite{montroll1965, metzler2004}. In this model, a searcher waits at its current location for a random time and then jumps a random distance chosen from some jump length probability density $f(y)$ in a uniform random direction. The searcher repeats these two steps indefinitely or until it reaches the target. For a finite mean waiting time $t_0\in(0,\infty)$ and a jump length density with the following slow power law decay,
\begin{align}\label{pl}
f(y)
\sim
 \frac{(l_0)^{2s}}{y^{1+2s}}\quad\text{as $y\to\infty$ for some $s\in(0,1)$ and lengthscale $l_0>0$},
\end{align}
the probability density ${{p}}(x,t)$ for the searcher position satisfies the following space-fractional Fokker-Planck equation in a certain scaling limit \cite{meerschaert2011},
%then the process is superdiffusive and is often called a L{\'e}vy flight \cite{metzler2004, dubkov2008}. 
%In a certain scaling limit, the probability density ${{p}}(x,t)$ for the position of a L{\'e}vy flight in $\R^{d}$ satisfies the space fractional Fokker-Planck equation \cite{meerschaert2011}.
%The probability density $p(x,t)$ for the position of a L{\'e}vy flight in $\R^{d}$ satisfies the space-fractional Fokker-Planck equation \cite{meerschaert2011},
\begin{align}\label{ffpe}
\frac{\partial}{\partial t}p
=-D_s(-\Delta)^{s}p,%\quad x\in\R^{d},\,t>0,
\end{align}
where
\begin{align}\label{Ds}
    D_s=(l_0)^{2s}/t_0>0
\end{align}
is the generalized diffusivity %(with dimension $[D_s]=(\textup{length})^{2s}/(\textup{time})$) 
and $(-\Delta)^{s}$ denotes the fractional Laplacian of order $s\in(0,1)$, defined by \cite{lischke2020}
\begin{align}\label{eq:frac-lap-def}
    (-\Delta)^s \varphi(x) = C_s\pvint_{-\infty}^\infty \frac{\varphi(x)-\varphi(y)}{|x-y|^{2s+d}}dy,\qquad C_s := \frac{4^s\Gamma(s+d/2)}{\pi^{d/2}|\Gamma(-s)|},
\end{align}
where P.V.\@ indicates the principal value and $\Gamma(\cdot)$ denotes the gamma function.  
Note that L{\'e}vy flights are often parameterized by their stability index $\alpha\in(0,2)$ \cite{palyulin2019}, which is simply twice the fractional order $s\in(0,1)$,
\begin{align*}
    \alpha=2s\in(0,2).
\end{align*}
Observe that \eqref{ffpe} is the diffusion equation describing Brownian motion if $s=1$. %Stochastic paths of L{\'e}vy flights differ qualitatively from Brownian paths due to the presence of very long jumps. 
% Indeed, the fractional equation in \eqref{ffpe} governing a L{\'e}vy flight can be derived from the continuous-time random walk model if the walker tends to take very large jumps (specifically, if the stochastic jump lengths have infinite variance) \cite{meerschaert2011}. Stochastic paths of L{\'e}vy flights can also be obtained via an appropriate random time change (subordination) of a Brownian path \cite{lawley2023super}.

L{\'e}vy flights are perhaps the most mathematically tractable model of superdiffusion, though analytical results for L{\'e}vy flights are scarce compared to their Brownian counterpart.
The mathematical analysis of hitting times of superdiffusive search processes has also been controversial. Indeed, the influential L{\'e}vy flight foraging hypothesis was based on the claimed theoretical optimality of a certain superdiffusive process involving heavy-tailed jumps as in \eqref{pl} with the ``inverse square" value $s=1/2$ \cite{viswanathan1999, viswanathan2008}, but this decades-old claim was recently shown to be false \cite{levernier2020, buldyrev2021, levernier2021reply}.

In this paper, we study FHTs of L{\'e}vy flights to small targets in one space dimension. Assuming the targets are much smaller than the typical distance between them, we apply the method of matched asymptotic expansions to the fractional differential equation describing the MFHT. The resulting asymptotic formulas reveal how FHTs depend on the fractional order $s\in(0,1)$, target size, target arrangement, and initial searcher location (or distribution of locations). We further determine the full probability distribution of the FHT for fractional orders $s\in(0,1/2]$ in the small target limit. We validate our results by comparison to both deterministic numerical solutions of the associated fractional differential equation and stochastic simulations.

To describe our results more precisely, let $X=\{X(t)\}_{t\ge0}$ be a one-dimensional, $(2s)$-stable L{\'e}vy flight for $s\in(0,1)$ with generalized diffusivity $D_s>0$ (i.e.\ the probability density that $X(t)=x$ satisfies \eqref{ffpe}) and periodic boundary conditions at $x=\pm l$. Since we can always re-scale space and time according to 
\begin{align}\label{rescale}
    x\to x/l,\quad t\to D_st/l^{2s}, 
\end{align}
we set $D_s=l=1$ without loss of generality. Suppose that the target set $\Omega_{\textup{target}}$ consists of $N\ge1$ targets in the interval $\Omega=(-1,1)\in\R$ centered at points $\{x_{1},\dots,x_{N}\}\in(-1,1)$ with radii $\{\eps l_{1},\dots,\eps l_{N}\}$, i.e.
\begin{align}\label{tll}
\Omega_{\text{target}}
=\cup_{i=1}^{N}(x_{i}-\eps l_{i},x_{i}+\eps l_{i}).
\end{align}
Here, $l_{1},\dots,l_{N}>0$ are $O(1)$ constants which allow the targets to differ in size. When the context is clear we denote by $|\cdot|$ the $2$-periodic extension of the absolute value on $(-1,1)$ so that $|a-b|$ denotes the minimum distance between $a$ and $b$ in the periodic domain $(-1,1$). Assume that $0<\eps\ll1$ and the targets are well-separated in the sense that $|x_{i}-x_{j}|\gg \eps$ for all $i,j\in\{1,\dots,N\}$ with $i\neq j$. Let $v(x)$ denote the MFHT to any of the $N$ targets starting from $x\in(-1,1)$, i.e.
\begin{align*}
v(x)
:=\E[\tau\,|\,X(0)=x],
\end{align*}
where $\tau$ is the FHT in \eqref{tau}. The function $v(x)$ satisfies
\begin{equation}\label{eq:hitting-time-pde}
	\begin{cases}
		(-\Delta)^s v(x) = 1, & x\in \Omega\setminus \Omega_{\text{target}}, \\
		v(x) = 0, & x\in\Omega_{\text{target}}, \\
		\text{$v(x)$ is $2$-periodic}.
	\end{cases}
\end{equation}
We obtain our results on the FHT by analyzing \eqref{eq:hitting-time-pde} in the limit $\eps\to0$.

We now state our results on the MFHT for the case of a single target of radius $\eps>0$ centered at $x_1=0$ (i.e.\ $N=l_1=1$). Note that our assumption of periodic boundary conditions means that this scenario is equivalent to a L{\'e}vy flight on all of $\R$ with a periodic array of targets separated by distance 2. For any fractional order $s\neq1/2$, the MFHT of a L{\'e}vy flight conditioned on starting at $x\in(-1,1)\setminus\{0\}$ is given by the following asymptotic formula for $0<\eps\ll1$
\begin{align}\label{s1}
v(x)
\sim\eps^{2s-1}\frac{2\mathfrak{a}_{s}}{\mathfrak{b}_{s}}
-2\mathfrak{a}_{s}R_{s}(0)
+2\mathfrak{a}_{s}\left(-|x|^{2s-1}+R_{s}(x)\right),
\end{align}
where
\begin{equation}\label{eq:def_of_a_and_b}
\mathfrak{a}_{s} := -2\pi^{-1}s\Gamma(-2s)\sin(\pi s),\quad
\mathfrak{b}_s:=\frac{\Gamma(1/2)}{\Gamma(3/2-s)\Gamma(s)},
\end{equation}
and $R_s$ is the regular part of the Green's function given explicitly in Proposition \ref{prop:greens-func}. 
If $s=1/2$, then this MFHT is
\begin{align}\label{s2}
v(x)\sim\log(2/\eps)\frac{2}{\pi}
-\frac{2}{\pi}R_{1/2}(0)
+\frac{2}{\pi}\big(\log|x|+R_{1/2}(x)\big).
\end{align}
If the L{\'e}vy flight searcher starts from a uniformly distributed position in the interval $(-1,1)$, then the average MFHT is
\begin{align}\label{unif0}
\frac{1}{2}\int_{-1}^{1}v(x)\,\dd x
\sim\displaystyle\begin{dcases}
\eps^{2s-1}2\mathfrak{a}_{s}/\mathfrak{b}_{s}
-R_{s}(0)2\mathfrak{a}_{s} & \text{if }s\neq1/2,\\
\log(2/\eps)2/\pi
-2R_{1/2}(0)/\pi
 & \text{if }s=1/2.
\end{dcases}
\end{align}

\begin{figure}[t!]
	\centering % <-- added
	\begin{subfigure}{0.33\textwidth}
		\includegraphics[width=\linewidth]{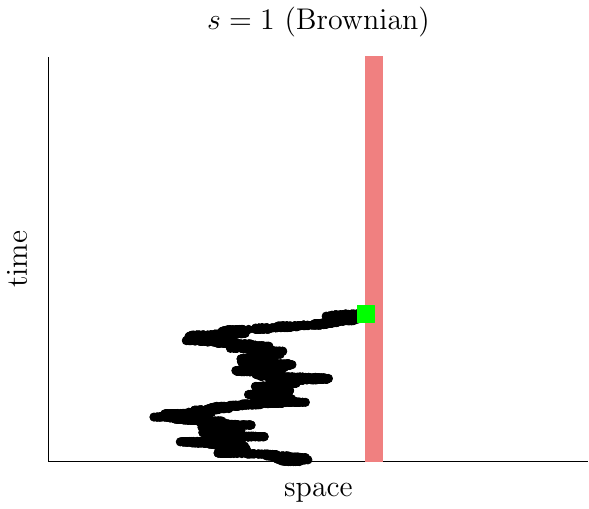}
	\end{subfigure}\hfil % <-- added
	\begin{subfigure}{0.33\textwidth}
		\includegraphics[width=\linewidth]{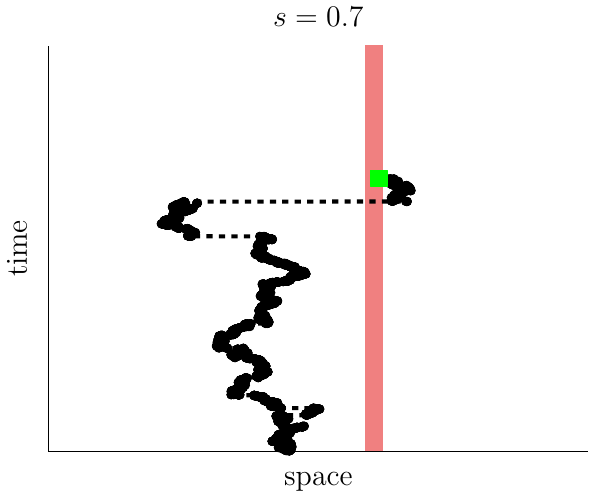}
	\end{subfigure}\hfil % <-- added
	\begin{subfigure}{0.33\textwidth}
		\includegraphics[width=\linewidth]{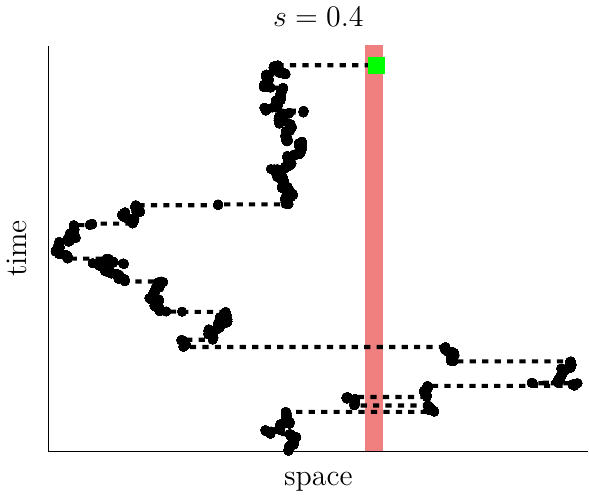}
	\end{subfigure}
	\caption{
 L{\'e}vy flight sample paths in one space dimension for (left) $s=1$ (i.e.\ Brownian motion), (middle) $s=0.7$, and (right) $s=0.4$. The solid black markers show positions of the L{\'e}vy flight. For $s<1$, the black dashed lines show the discontinuous jumps of the L{\'e}vy flight, which become larger for smaller values of $s\in(0,1)$ and allow the L{\'e}vy flight to jump across the target (red region).}\label{figschem1d}
\end{figure}

These results show an analog between Brownian search in dimensions $d\ge1$ and L{\'e}vy search in dimension $d=1$ with fractional order $s\in(0,1)$.
%In the small target limit, we find that the mean FHT diverges for any fractional order $s\le1/2$. 
Specifically, \eqref{s1}-\eqref{unif0} imply
\begin{align}\label{super}
\E[\tau]
=\begin{cases}
O(1) &\text{if }s\in(1/2,1],\\
O(\log (1/\eps)) &\text{if }s=1/2,\\
O(\eps^{2s-1}) &\text{if }s\in(0,1/2).
\end{cases}
\end{align}
%Comparing \eqref{diff} to \eqref{super} shows that FHTs of Brownian motion in dimensions (i) $d=1$, (ii) $d=2$, and (iii) $d\ge3$ correspond respectively to FHTs of L{\'e}vy flights in dimension $d=1$ with fractional order (i) $s\in(1/2,1]$, (ii) $s=1/2$, and 
Comparing \eqref{diff} to \eqref{super} shows that FHTs of Brownian motion in different dimensions diverge similarly to FHTs of L{\'e}vy flights in one dimension with different fractional orders. As in the case of Brownian motion in \eqref{diff}, the different regimes in \eqref{super} stem from differences in recurrence versus transience, which manifests in our analysis as different far-field behavior of the inner solutions used in our matched asymptotics. FHTs of L{\'e}vy flights in one dimension can diverge because the stochastic paths of L{\'e}vy flights are discontinuous. Hence, in contrast to Brownian motion, L{\'e}vy flights may jump across a target without actually hitting it in a phenomenon termed a ``leapover'' \cite{koren2007, koren2007first, palyulin2019, wardak2020} (see Figure~\ref{figschem1d} for an illustration).

Our analysis allows us to identify the value of $s\in(0,1]$ which minimizes the MFHT. We find that this optimal value (denoted $s_{\textup{opt}}$) grows continuously from $s_{\textup{opt}}\approx0$ up to $s_{\textup{opt}}\approx1$ (i.e.\ Brownian search) as the target density grows relative to the lengthscale $l_0$ in \eqref{pl}-\eqref{Ds}. In particular, we show that the value $s=1/2$ (corresponding to stability index $\alpha=2s=1$, i.e.\ inverse square L\'evy search) is optimal in only very specific circumstances.

The rest of the paper is organized as follows. %In section~\ref{mainresults}, we formulate the model precisely and summarize our main results. 
In Section~\ref{sectionfht}, we analyze the mean and full probability distribution of the FHT. In Section~\ref{sec:numerical}, we compare our asymptotic results to numerical solutions of the associated fractional equations and stochastic simulations. In Section~\ref{sectionopt}, we address the question of the fractional order $s\in(0,1]$ that minimizes the MFHT. We conclude by summarizing our results and discussing related work. An appendix collects some more technical aspects of the numerical implementation in Section~\ref{sec:numerical}.

\section{Asymptotic Analysis of the MFHT}\label{sectionfht}

The method of matched asymptotic expansions (MMAE) has been an invaluable tool in the analysis of narrow capture and escape problems for pure diffusion processes since its introduction in \cite{ward10,ward10b}. Broadly speaking, the MMAE proceeds by formulating inner- and outer-problems whose solutions can be expressed in terms of a canonical ``electrified disk'' solution and an appropriately weighted sum of Green's functions respectively. Combining a solvability condition for the outer-problem together with matching conditions between the inner- and outer-solutions yields a linear system with which all remaining unknowns arising in the asymptotic analysis can be determined. In this section we adapt the MMAE to derive an asymptotic expansion for the MFHT satisfying the fractional differential equation \eqref{eq:hitting-time-pde}. We show how the MMAE in this fractional setting synthesizes the analysis of the standard narrow escape problem in dimensions $d=2$ and $d=3$. In addition we introduce a fractional counterpart to the classical electrified disk problem, as well as a $2$-periodic fractional Green's function.  

We begin our asymptotic analysis of the MFHT by seeking an outer asymptotic expansion of the form
\begin{subequations}
\begin{equation}\label{eq:outer-def}
    v(x) \sim \ep^{2s-1}v_0^\ep(x),
\end{equation}
valid for values of $x$ that are sufficiently far from all targets in the sense that $|x-x_i|\gg \ep$ for all $i=1,...,N$. In addition for each $i=1,...,N$ we seek an inner asymptotic expansion of the form
\begin{equation}\label{eq:inner-def}
    v(x_i+\ep X) \sim V_{i}^\ep(X),
\end{equation}
\end{subequations}
valid for values of $x=x_i+\ep X$ sufficiently close to the $i^\text{th}$ target in the sense that $X=O(1)$.

It is here convenient to recall two equivalent definitions of the fractional Laplacian given by \eqref{eq:frac-lap-def} when restricted to $2$-periodic functions. Specifically, if we let $\varphi(x)$ be an arbitrary $2$-periodic function then
\begin{subequations}\label{eq:frac-lap-periodic-kernel}
\begin{equation}
(-\Delta)^s \varphi(x) = C_s  \pvint_{-1}^1 K_s(x-y)(\varphi(x)-\varphi(y))dy,
\end{equation}
where
\begin{equation}
    K_s(z):= \sum_{n\in\mathbb{Z}} \frac{1}{|z+2n|^{2s+1}},
\end{equation}
\end{subequations}
and where $\mathbb{Z}$ denotes the set of all integers.  This expression is conveniently chosen to determine appropriate inner problems. Moreover, it can be shown (see, for example, Eq.~(2.53) in \cite{abatangelo_2019}) that the restriction of the fractional Laplacian defined by \eqref{eq:frac-lap-def} to $2$-periodic functions coincides with the spectral fractional Laplacian defined by
\begin{equation}\label{eq:frac-lap-spectral}
    (-\Delta)^s \varphi(x) = \sum_{n=\mathbb{Z}\setminus\{0\}} |n\pi|^{2s}\varphi_n e^{in\pi x},\qquad \varphi_n := \frac{1}{2}\int_{-1}^1 e^{-in\pi x}\varphi(x)dx.
\end{equation}
This formulation proves to be useful when considering global quantities, such as the relevant periodic fractional Green's function.

In order to state our main result for this section we first define the scalars
\begin{subequations}\label{eq:scalar-vector-matrix-definitions-00}
\begin{equation}
    \nu_i^\ep := -\frac{1}{\log(\varepsilon l_i / 2)},\qquad \bar{\nu}^\ep := \frac{1}{N}\sum_{i=1}^N \nu_i^\ep,\qquad \bar{l}_s := \frac{1}{N}\sum_{i=1}^N l_i^{1-2s},
\end{equation}
as well as the $N$-dimensional vectors
\begin{equation}
    \pmb{l}_s := \begin{pmatrix} l_1^{1-2s} \\ \vdots \\ l_N^{1-2s} \end{pmatrix},\quad \pmb{\nu}^\ep := \begin{pmatrix} \nu_1^\ep \\ \vdots \\ \nu_N^\ep \end{pmatrix},\quad \pmb{e}_N := \begin{pmatrix} 1 \\ \vdots \\ 1 \end{pmatrix}.
\end{equation}
In addition, we define the $N\times N$ diagonal matrices
\begin{equation}
    \mathcal{L}_s := \diag(l_1^{1-2s},...,l_N^{1-2s}),\qquad \mathcal{N}^\ep := \diag(\nu_1^\ep,...,\nu_N^\ep),
\end{equation}
as well as the $N\times N$ \textit{Green's matrix} $\mathcal{G}_s$ whose entries are given by
\begin{equation}
    (\mathcal{G}_s)_{ij} = \begin{cases} R_s(0), & i=j, \\ R_s(x_i-x_j) + H_s(x_i-x_j), & i\neq j,\end{cases}
\end{equation}
\end{subequations}
where $R_s$ is the regular part of the Green's function defined in Proposition \ref{prop:greens-func}, and $H_s(z)$ is the singular part with $H_s(z):=-|z|^{2s-1}$ for $s\neq1/2$ and $H_s(z):=\log|z|$ for $s=1/2$. Our main asymptotic result for the hitting time is given below.

\begin{principal_result}\label{princ:hitting-time}
Let $\ep\ll 1$, let $l_1,...,l_N=O(1)$, and suppose that $-1\leq x_1<...<x_N<1$ are well separated in the sense that $|x_i-x_j|\gg O(\ep)$ for all $i\neq j$. For any $0<s<1$ define
\begin{subequations}
\begin{equation}\label{eq:chi-def}
    \chi^\ep := \begin{dcases} \frac{1}{N\bar{l}_s}\left(\frac{2\mathfrak{a}_s}{\mathfrak{b}_s} - \ep^{1-2s}\mathfrak{b}_s \pmb{l}_s^T\mathcal{G}_s\mathcal{L}_s\pmb{B}^\ep \right), & s\neq 1/2, \\ 
    \frac{2}{\pi N \bar{\nu}^\ep}\left( 1 - \frac{\pi}{2}(\pmb{\nu}^\ep)^T\mathcal{G}\pmb{B}^\ep\right) ,& s=1/2,\end{dcases}
\end{equation}
where $\mathfrak{a}_s$ and $\mathfrak{b}_s$ are given by \eqref{eq:def_of_a_and_b}, and where the $N$-dimensional vector $\pmb{B}^\ep=(B_1^\ep,...,B_N^\ep)^T$ is found by solving the linear system
\begin{equation}\label{eq:B-sys}
    \begin{dcases}
        \left( \mathcal{I}_N - \ep^{1-2s}\mathfrak{b}_s\left( \mathcal{I}_N - \frac{1}{N\bar{l}_s}\pmb{e}_N\pmb{l}_s^T \right)\mathcal{G}_s\mathcal{L}_s \right) \pmb{B}^\ep = \frac{2\mathfrak{a}_s}{N\bar{l}_s \mathfrak{b}_s}\pmb{e}_N,& s\neq 1/2, \\
        \left(\mathcal{I}_N - \mathcal{N}^\ep\left( \mathcal{I}_N - \frac{1}{N\bar{\nu}^\ep}\pmb{e}_N(\pmb{\nu}^\ep)^T  \right)\mathcal{G}_{1/2}\right)\pmb{B}^\ep = \frac{2}{\pi  N \bar{\nu}^\ep}\pmb{\nu}^\ep,& s=1/2,
    \end{dcases}
\end{equation}
where $\mathcal{I}_N$ is the $N\times N$ identity matrix. Then, an asymptotic expression for the MFHT satisfying \eqref{eq:hitting-time-pde} for $|x-x_i|\gg \ep$ for all $i=1,...,N$ is given by
\begin{equation}\label{eq:v-asy-sol}
v(x) \sim \ep^{2s-1}\chi^\ep + \begin{dcases} \mathfrak{b}_s\sum_{j=1}^{N} l_j^{1-2s}B_j^\ep (-|x-x_j|^{2s-1} + R_s(x-x_j)), & s\neq 1/2, \\
\sum_{j=1}^N B_j^\ep \left(\log|x-x_j| +  R_{1/2}(x-x_j)  \right),& s=1/2,\end{dcases}
\end{equation}
\end{subequations}
where $R_s(x)$ is the regular part of the Green's function found in Proposition \ref{prop:greens-func}.
\end{principal_result}

The remainder of this section is organized as follows. In Section \ref{subsec:electrified-disk} we first establish some key properties of a fractional counterpart to the classical electrified disk problem. This is followed by a discussion of a certain $2$-periodic fractional Green's function in Section \ref{subsec:greens-func}. In Section \ref{subsec:hitting-time-matched-asymptotics} we then proceed with applying the MMAE to derive Principal Result \ref{princ:hitting-time}. Finally, in Section \ref{subsec:higher-moments} we show that, to leading order, the FHT $\tau$ is exponentially distributed for $s\in(0,1/2].$

\subsection{The Fractional Electrified Disk Problem}\label{subsec:electrified-disk}

The fractional counterpart to the electrified disk problem in standard narrow capture problems is the problem
\begin{equation}\label{eq:inner-problem-Ws}
	\begin{cases}
		(-\Delta)^s W_s(X) = 0,& |X|>1, \\
		W_s(X) = 1, & |X|<1.
	\end{cases}
\end{equation}
The function $W_s(X)$ is the probability that a L\'evy flight starting at $X\in\mathbb{R}$ eventually hits the ball $(-1,1)$. With this probabilistic interpretation, one readily obtains the following formula for $W_s(X)$ when $s<1/2$ (see Corollary 2 in \cite{blumenthal_1961})
\begin{equation}\label{eq:Ws-blumenthal}
	W_s(X) = \frac{\sqrt{\pi}}{\Gamma(\tfrac{1}{2} - s)\Gamma(s)}\int_{X^2-1}^\infty\frac{u^{s-1}}{\sqrt{u+1}}du.
\end{equation}
We proceed to derive an explicit expression for $W_s(X)$ valid for all $s\in(0,1)$. Specifically, we deploy a Kelvin transform and fractional Poisson formula for $s\neq 1/2$, and standard complex analysis tools for $s=1/2$. The main result is summarized in the following proposition.

\begin{proposition}\label{prop:electrified-disk}
	The fractional electrified disk problem \eqref{eq:inner-problem-Ws} admits the following non-constant solution
	\begin{subequations}
		\begin{equation}\label{eq:Ws-sol}
			W_s(X) = \begin{cases}
				\frac{\sqrt{\pi}}{\Gamma(s)\Gamma(\tfrac{3}{2} - s)}|X|^{2s-1} \left(1 - \frac{1}{X^2}\right)^s\, _2F_1\left(1,\frac{1}{2};\frac{3}{2} -s; \frac{1}{X^2}\right), & s\neq 1/2, \\
				1-\log(X+\sqrt{X^2-1}), & s=1/2,
			\end{cases}\quad |X|>1,
		\end{equation}
		with $W_s(X)=1$ for $|X|\leq 1$. Moreover, this solution has the far-field behaviour
		\begin{equation}\label{eq:Ws-far-field}
			W_s(X)\sim \begin{cases}
				\mathfrak{b}_s|X|^{2s-1} + O(|X|^{2s-3}),& s\neq 1/2, \\
				-\log(2|X|) + 1 + O(X^{-2}), & s=1/2,
			\end{cases}\qquad \text{as}\quad|X|\rightarrow\infty,
		\end{equation}
	\end{subequations}
	where $\mathfrak{b}_s$ is given by \eqref{eq:def_of_a_and_b}.
\end{proposition}

Starting with the $s\neq 1/2$ case, we first transform \eqref{eq:inner-problem-Ws} into the more commonly considered fractional problem with extended Dirichlet boundary conditions posed outside of $(-1,1)$. Specifically, we first use the Kelvin transform
\begin{subequations}
	\begin{equation}\label{eq:kelvin-tx-def}
		\overline{X} = 1/X,\qquad \overline{W}_s(\overline{X}) = |\overline{X}|^{2s-1}W_s\left(1/\overline{X}\right),
	\end{equation}
	in terms of which we readily calculate (see, for example, Proposition A.1 in \cite{oton_2014})
	\begin{equation}\label{eq:kelvin-tx-frac-lap}
		(-\Delta)^s\overline{W}_s(\overline{X}) = |X|^{2s+1}(-\Delta)^sW_c(X).
	\end{equation}
\end{subequations}
In particular we find that $\overline{W}_s(\overline{X})$ solves
\begin{equation}\label{eq:inner-problem-Ws-bar}
	\begin{cases}
		(-\Delta)^s \overline{W}_s(\overline{X}) = 0,& |\overline{X}|<1, \\
		\overline{W}_s(\overline{X}) = |\overline{X}|^{2s-1}, & |\overline{X}|>1.
	\end{cases}
\end{equation}

Notice that the inhomogeneous term $g(\overline{X}) = |\overline{X}|^{2s-1}$ for $|\overline{X}|>1$ in \eqref{eq:inner-problem-Ws-bar} can be extended to $\mathbb{R}$ in such a way that $g\in L^1_{\text{loc}}(\mathbb{R})\cap C(\mathbb{R})$ and 
\begin{equation*}
	\int_{\mathbb{R}}\frac{|g(\overline{X})|}{1+|\overline{X}|^{1+2s}} d\overline{X} <\infty .
\end{equation*}
It then follows that the unique \textit{continuous} solution to \eqref{eq:inner-problem-Ws-bar} is given by (see Theorem 2.10 in \cite{bucur_2016})
\begin{equation*}
	\overline{W}_s(\overline{X}) = \begin{cases} \int_{|Y|>1}P_s(Y,\overline{X})|Y|^{2s-1} dY, & |\overline{X}|<1, \\ |\overline{X}|^{2s-1}, & |\overline{X}| > 1,	\end{cases}
\end{equation*}
where $P_s(y,x)$ is the fractional Poisson kernel given by
\begin{equation*}
	P_s(y,x) := p_s\left(\frac{1-x^2}{y^2-1} \right)^s \frac{1}{|x-y|},\qquad p_s := \pi^{-1}\sin(\pi s) = \frac{1}{\Gamma(s)\Gamma(1-s)}.
\end{equation*}
Reverting to the original variables we therefore obtain the integral representation
\begin{equation*}
\begin{split}
	W_s(X) &= p_s \int_1^\infty \left( \tfrac{X^2-1}{1-1/Y^2}\right)^s  \tfrac{2|X|}{(XY)^2-1}dY \\
 &= p_s |X|^{2s-1}(1-\tfrac{1}{X^2})^{s}\int_0^\infty \frac{(z+1)^{s-\tfrac{1}{2}}}{z^s (z+1-\tfrac{1}{X^2})}dz,
\end{split}
\end{equation*}
where the first equality follows by combining the $Y\in(-\infty,-1)$ and $Y\in(1,\infty)$ contributions, and the second from the change of variables $Y = \sqrt{z+1}$. Using the integral representation of the Gaussian Hypergeometric function (see Eq.\@ 15.6.1 in \cite{NIST:DLMF}) we immediately obtain \eqref{eq:Ws-sol}. The far-field behaviour \eqref{eq:Ws-far-field} of $W_s(X)$ is likewise immediately obtained by noting that (see Eq. 15.2.1 in \cite{NIST:DLMF})
\begin{equation*}
	_2F_1(1,\tfrac{1}{2};\tfrac{3}{2}-s;z) = 1 + \frac{z}{3-2s} + \frac{3z^2}{4s^2-16s+15} + O(z^3),\qquad |z|\ll 1.
\end{equation*}

\begin{remark}
	The equivalence of \eqref{eq:Ws-sol} and \eqref{eq:Ws-blumenthal} is readily verified using  properties of the Gaussian Hypergeometric function. Specifically, we first recast the integral in \eqref{eq:Ws-blumenthal} in terms of the Gaussian Hypergeometric function using the change of variables $u= X^2 (z+1)-1$. Equivalence with \eqref{eq:Ws-sol} is then verified by first using Euler's transformation $_2F_1(a,b;c;z) = (1-z)^{c-a-b}\, _2F_1(c-a,c-b;c;z)$ and then using the symmetry property $_2F_1(a,b;c;z)=\, _2F_1(b,a;c;z)$.
\end{remark}

We consider next the case $s=1/2$ for which the previous calculations yield $W_s(X)\equiv 1$. Indeed, it is easy to see that $\overline{W}_s(\overline{X}) \equiv 1$ is the unique continuous solution to \eqref{eq:inner-problem-Ws-bar} when $s=1/2$. To find a \textit{non-constant} solution to \eqref{eq:inner-problem-Ws} we instead consider the extended problem in the two-dimensional upper half-space. Specifically, we seek a non-constant solution $\widetilde{W}(X,Y)$ to
\begin{equation}
	\begin{cases}
		\frac{\partial^2 \widetilde{W}}{\partial X^2} + \frac{\partial^2 \widetilde{W}}{\partial Y^2} = 0, & -\infty<X<\infty,\, Y>0, \\
		\widetilde{W} = 1, & |X| < 1,\, Y=0, \\
		\frac{\partial\widetilde{W}}{\partial Y} = 0, & |X|>1,\, Y=0,
	\end{cases}
\end{equation}
in terms of which $W_{s=1/2}(X) = \widetilde{W}(X,0)$ (see \cite{caffarelli_2007} for additional details on the extension property of the fractional Laplacian). Such a non-constant solution must have logarithmic growth as $X^2+Y^2\rightarrow \infty$ and is given by
\begin{equation*}
	\widetilde{W}(X,Y) = 1 + \text{Im}\{\arcsin(X+iY)\},
\end{equation*}
where $\text{Im}(z)$ denotes the imaginary part of $z\in\mathbb{C}$. Setting $Y=0$ and considering only values of $|X|>1$ we readily obtain \eqref{eq:Ws-sol} from which the  far-field behaviour \eqref{eq:Ws-far-field} immediately follows.

\subsection{The Periodic Fractional Green's Function}\label{subsec:greens-func}

The second quantity we need to apply the MMAE is the periodic fractional Green's function $G_s(x)$ satisfying 
\begin{equation}\label{eq:greens-pde}
	\begin{cases}
		(-\Delta)^s G_s(x) = \frac{1}{2} - \delta(x), & -1<x<1, \\
		G_s(x+2) = G(x), & -\infty<x<\infty, \\
		\int_{-1}^1 G_s(x)dx = 0.
	\end{cases}
\end{equation}
Using the spectral definition of the fractional Laplacian \eqref{eq:frac-lap-spectral} it is straightforward to see that
\begin{equation}\label{eq:greens-series}
	G_s(x) = - \sum_{n=1}^\infty \frac{\cos n\pi x}{(n\pi)^{2s}}.
\end{equation}
We readily see that $G_s(x)$ diverges as $x\rightarrow 0$ for $s\leq 1/2$. The following proposition extracts this singular behaviour and decomposes $G_s(x)$ into a \textit{singular} part and a \textit{regular} part.
\begin{proposition}\label{prop:greens-func}
	The periodic fractional Green's function $G_s(x)$ satisfying \eqref{eq:greens-pde} is given by 
	\begin{subequations}
		\begin{equation}\label{eq:greens-func-decomposed}
			G_s(x) = \begin{cases} -\mathfrak{a}_{s}|x|^{2s-1} + \mathfrak{a}_sR_s(x) , & s\neq 1/2, \\ \pi^{-1}\log|x| + \pi^{-1}R_{1/2}(x), & s=1/2, \end{cases}
		\end{equation}
		where $\mathfrak{a}_{s}$ is given by \eqref{eq:def_of_a_and_b}. When $s\neq 1/2$ the regular part $R_s(x)$ admits the following rapidly converging series
		\begin{equation}\label{eq:greens-func-reg-s-not-1/2}
			\begin{split}
				R_s(x) = & \tfrac{1}{2s} - \tfrac{2s-1}{6} + \tfrac{7}{15}\tfrac{(2s-1)(2s-2)(2s-3)}{24}  + \left(\tfrac{2s-1}{2} - \tfrac{(2s-1)(2s-2)(2s-3)}{12}\right) |x|^2 \\
				& + \tfrac{(2s-1)(2s-2)(2s-3)}{24}|x|^4 + 2(2s-1)\cdots(2s-5)\sum_{n=1}^\infty \tfrac{a_{2s,n}}{(\pi n)^{2s}}\cos(\pi n x),
			\end{split}
		\end{equation}
		where $a_{2s,n} = \int_{\pi n}^\infty x^{2s-6}\sin x dx$. On the other hand, when $s=1/2$ the regular part has the series expansion
		\begin{equation}\label{eq:greens-func-reg-s=1/2}
			R_{1/2}(x) = 1 + 2\sum_{n=1}^\infty \left( \Si(n\pi) - \frac{\pi}{2}\right) \frac{\cos n\pi x}{n\pi},
		\end{equation}
		where $\Si(z) = \int_0^z t^{-1}\sin(t)dt$ denotes the usual sine integral.
	\end{subequations}
\end{proposition}

The calculation of $G_s(x)$ in the case $s\neq 1/2$ follows from computing Fourier series of $|x|^2$, $|x|^4$, and $|x|^{2s-1}$ and can be found in Appendix A of \cite{gomez_2023}. The case $s=1/2$ follows similarly, but this time only the Fourier series of $\log|x|$ is needed.

For the subsequent asymptotic analysis the most important part of $G_s(x)$ in \eqref{eq:greens-func-decomposed} is the \textit{singular} behaviour which takes the form of an algebraic singularity for $s<1/2$, a logarithmic singularity for $s=1/2$, and a bounded fractional cusp for $s>1/2$. The series expansions for the regular part appearing in \eqref{eq:greens-func-reg-s-not-1/2} and \eqref{eq:greens-func-reg-s=1/2} on the other hand are computationally useful due to their fast convergence.

\subsection{Matched Asymptotic Expansions}\label{subsec:hitting-time-matched-asymptotics}

Let $x=x_i+\ep X$ and substitute the inner expansion \eqref{eq:inner-def} into \eqref{eq:hitting-time-pde} so that using \eqref{eq:frac-lap-periodic-kernel} for the fractional Laplacian we obtain
\begin{equation}\label{eq:inner-eq-raw}
	\ep C_s\pvint_{-1/\ep}^{1/\ep}\sum_{n\in\mathbb{Z}}\frac{V_i^\ep(X)-V_i^\ep(Y)}{|2n+\ep(X-Y)|^{2s+1}}dY +\hot= 1,
\end{equation}
where $\hot$ denotes higher-order-terms. The $n=0$ term dominates all other terms in the left-hand-side, and moreover we will also assume that it dominates the right-hand-side by assuming that $V_i^\ep \gg \ep^{2s}$ for all $X=O(1)$. Further approximating the integral on the left-hand-side by replacing $\pm1/\ep$ with $\pm\infty$ we thus obtain the inner problem
\begin{equation}
	\begin{cases}
		(-\Delta)^s V_i^\ep (X) = 0, & |X|>l_i,\\
		V_i^\ep(X) = 0, & |X|\leq l_i, \\
	\end{cases}
\end{equation}
where the limiting behaviour of $V_i^\ep(X)$ as $|X|\rightarrow\infty$ will be found by matching with the limiting behaviour of the outer solution as $x\rightarrow x_i$ for each $i=1,...,N$.

In light of Proposition \ref{prop:electrified-disk} we seek, for each $i=1,...,N$, a non-constant inner solution of the form
\begin{equation}\label{eq:inner-sol-0}
	V_i^\ep(X) = \ep^{2s-1}B_i^\ep\left(1 - W_s(X/l_i)\right),
\end{equation}
where $B_i^\ep$ is some $\ep$-dependent constant that remains to be determined. From Proposition \ref{prop:electrified-disk} we then have the far-field behaviour
\begin{equation*}
	V_i^\ep(X) \sim \begin{cases} \ep^{2s-1}B_i^\ep\left(1 - \mathfrak{b}_sl_i^{1-2s}|X|^{2s-1} + O(|X|^{2s-3})   \right), & s\neq 1/2, \\
 B_i^\ep\left(\log(2|X/l_i|) + O(|X|^{-2}) \right),& s=1/2,
	\end{cases}\quad \text{as }\, |X|\rightarrow \infty.
\end{equation*}
The far-field behaviour of $V_i^\ep(X)$ must coincide with the limiting behaviour of the outer solution $v^\ep_0(x)$ as $x\rightarrow x_i$. Specifically, writing $X=\ep^{-1}(x-x_i)$ we obtain the \textit{matching condition} as $|x-x_i|\rightarrow 0$,
\begin{equation}\label{eq:matching-0}
	v^\ep_0(x) \sim \begin{cases} B_i^\ep\left(1 - \mathfrak{b}_s\ep^{1-2s}l_i^{1-2s}|x-x_i|^{2s-1} + O(\ep^{3-2s})\right), & s\neq 1/2, \\ B_i^\ep \left( \log|x-x_i|   +1/\nu_i^\ep + O(\ep^2)\right),& s=1/2.\end{cases}
\end{equation}

Given the singular term $|x-x_i|^{2s-1}$ in the limiting behaviour \eqref{eq:matching-0} we find that $v^\ep_0(x)$ is the $2$-periodic function satisfying
\begin{equation}\label{eq:outer-eq-0}
	(-\Delta)^s v_0^\ep(x) = \begin{cases}\ep^{1-2s} - \ep^{1-2s}\mathfrak{a}_s^{-1}\mathfrak{b}_s\sum_{j=1}^N l_j^{1-2s}B_j^\ep \delta(x-x_j), & s\neq 1/2, \\
  1 - \pi\sum_{j=1}^N B_j^\ep\delta(x-x_j),& s=1/2,\end{cases}
\end{equation}
Since this problem is posed on the whole (periodic) interval $-1<x<1$, we can now use the spectral definition \eqref{eq:frac-lap-spectral} for the fractional Laplacian so that by integrating \eqref{eq:outer-eq-0} over the domain we obtain the solvability conditions
\begin{equation}\label{eq:solvability-0}
	\mathfrak{a}_s^{-1}\mathfrak{b}_s\sum_{j=1}^N l_j^{1-2s}B_j^\ep  = 2,\qquad \sum_{j=1}^N B_j^\ep = \frac{2}{\pi},
\end{equation}
for $s\neq1/2$ and $s=1/2$ respectively. Provided this condition is satisfied, we can then write $v^\ep_0(x)$ in terms of the periodic fractional Green's function found in Proposition \ref{prop:greens-func} as
\begin{equation}\label{eq:outer-0}
	v^\ep_0(x) = \chi^\ep + \begin{cases} \ep^{1-2s}\mathfrak{b}_s\sum_{j=1}^{N} l_j^{1-2s}B_j^\ep (-|x-x_j|^{2s-1} + R_s(x-x_j)),& s\neq 1/2, \\ \sum_{j=1}^N B_j^\ep\left( \log|x-x_j| + R_{1/2}(x-x_j)\right),& s=1/2,\end{cases}
\end{equation}
where $\chi^\ep$ is an undetermined constant.

The asymptotic analysis has thus far yielded an expression for the outer solution in terms of the $N+1$ unknown quantities $B_1^\ep,...,B_N^\ep$ and $\chi^\ep$. The solvability condition \eqref{eq:solvability-0} yields one equation in these $N+1$ unknowns. By revisiting the matching condition \eqref{eq:matching-0} we obtain the remaining $N$ equations with which all $N+1$ unknowns can be uniquely determined. Specifically, substituting the asymptotic expansion of \eqref{eq:outer-0} as $x\rightarrow x_i$ into the left-hand-side of \eqref{eq:matching-0} gives the matching condition
\begin{equation*}
	\ep^{1-2s}\mathfrak{b}_s l_i^{1-2s}B_i^\ep R_s(0) + \ep^{1-2s}\mathfrak{b}_s\sum_{j\neq i} l_j^{1-2s}B_j^\ep (-|x_i-x_j|^{2s-1} + R_s(x_i-x_j) )+ \chi^\ep = B_i^\ep,
\end{equation*}
when $s\neq 1/2$, and
\begin{equation*}
	B_i^\ep R_{1/2}(0) + \sum_{j\neq i}B_j^\ep\left(\log|x_i-x_j| + R_{1/2}(x_i-x_j)\right) + \chi^\ep = B_i^\ep/\nu_i^\ep,
\end{equation*}
when $s=1/2$ for each $i=1,...N$. In light of the definitions \eqref{eq:scalar-vector-matrix-definitions-00} we can rewrite the solvability and matching conditions in vector notation as
\begin{equation*}
\begin{cases}
    \pmb{l}_s^T\pmb{B}^\ep = \frac{2\mathfrak{a}_s}{\mathfrak{b}_s},\quad \pmb{B}^\ep - \ep^{1-2s}\mathfrak{b}_s \mathcal{G}_s\mathcal{L}_s\pmb{B}^\ep = \chi^\ep \pmb{e}_N, & s\neq 1/2, \\
    \pmb{e}_N^T\pmb{B}^\ep = \frac{2}{\pi},\quad \pmb{B}^\ep - \mathcal{N}^\ep \mathcal{G}_{1/2}\pmb{B}^\ep = \chi^\ep\pmb{\nu}^\ep, & s=1/2.    
\end{cases}
\end{equation*}
Left-multiplying the matching condition in the $s\neq 1/2$ (respectively $s=1/2$) case by $\pmb{l}_s^T$ (respectively $\pmb{e}_N^T$) and using the solvability condition yields the expression for $\chi^\ep$ found in \eqref{eq:chi-def}. Substituting this expression for $\chi^\ep$ back into the matching condition then gives the linear system \eqref{eq:B-sys}.

We claim that the solution $\pmb{B}^\ep$ to \eqref{eq:B-sys} is $O(1)$ for all $s\in(0,1)$. Indeed, when $s<1/2$ we readily obtain the expansion
\begin{equation*}
	\pmb{B}^\ep = \frac{2\mathfrak{a}_s}{N\bar{l}_s\mathfrak{b}_s}\sum_{q=0}^{\infty}\ep^{q(1-2s)}\mathcal{J}_s^q \pmb{e}_N,\qquad \mathcal{J}_s := \mathfrak{b}_s\left( \mathcal{I}_N - \frac{1}{N\bar{l}_s}\pmb{e}_N\pmb{l}_s^T \right)\mathcal{G}_s\mathcal{L}_s.
\end{equation*}
Similarly, when $s=1/2$ we obtain an expansion in powers of $\nu_1^\ep,...,\nu_N^\ep$ starting with an $O(1)$ term since $\pmb{\nu}^\ep / \bar{\nu}^\ep = O(1)$. When $s>1/2$ we must proceed by imposing a solvability condition. Specifically, assuming that $\mathcal{G}_s$ is invertible we find that the kernel of $\mathcal{J}_s$ is one-dimensional and spanned by  $\pmb{\xi}_s = \mathcal{L}_s^{-1}\mathcal{G}_s^{-1}\pmb{e}_N$. Seeking an expansion of the form $\pmb{B}^\ep = \pmb{B}_0 + \ep^{2s-1}\pmb{B}_1 + \cdots$ and imposing a solvability condition for the $\pmb{B}_1$ equation yields
\begin{equation*}
    \pmb{B}^\ep = \gamma_0 \pmb{\xi}_s + O(\ep^{2s-1}),\qquad \gamma_0 = \frac{2\mathfrak{a}_s}{N\bar{l}_s \mathfrak{b}_s}\frac{\pmb{l}_s^T\pmb{e}_N}{\pmb{l}_s^T\pmb{\xi}_s}.
\end{equation*}

The preceding discussion implies that our asymptotic expansion is consistent with the assumption $V_i^\ep(X)\gg \ep^{2s}$ that we made to neglect the inhomogeneous term on the right-hand-side of \eqref{eq:inner-eq-raw}.

\begin{remark}
    Since $\pmb{B}^\ep=O(1)$ for all $0<s<1$, we deduce from \eqref{eq:chi-def} that $\chi^\ep = O(1)$ for $s\leq 1/2$ whereas $\chi^\ep=O(\ep^{1-2s})$ for $1/2<s<1$. Hence \eqref{eq:v-asy-sol} implies that to leading order the MFHT in the outer region is spatially constant for $s\leq 1/2$ whereas it is spatially variable for $1/2<s<1$.
\end{remark}

\begin{remark}
    If the target configuration is symmetric, in the sense that $l_1=...=l_N=l$ and adjacent targets are equidistant, then $\nu_1=...=\nu_N=\nu$, the Green's matrix $\mathcal{G}_s$ is circulant, $\mathcal{L}_s=l\mathcal{I}_N$, and $\mathcal{N}^\ep=\nu\mathcal{I}_N$. The solution to \eqref{eq:B-sys} is then explicitly given by $\pmb{B}^\ep = \tfrac{2\mathfrak{a}_s}{Nl\mathfrak{b}_s}\pmb{e}_N$ and $\pmb{B}^\ep = \tfrac{2}{\pi N }\pmb{e}_N$ for $s\neq 1/2$ and $s=1/2$ respectively. Moreover, it suffices to consider symmetric configurations for only $N=1$ since the case $N>1$ can be obtained by a simple spatial rescaling.
\end{remark}

\subsection{Probability distribution for \texorpdfstring{$s\in(0,1/2]$}{0<s<1/2}}\label{subsec:higher-moments}

We now extend the preceding analysis of the MFHT to obtain the full probability distribution of the FHT in the limit $\ep\to0$ for $s\in(0,1/2]$. The $m$th moment of the FHT,
\begin{align*}
v_m(x)
:=\E[\tau^m\,|\,X(0)=x],\quad m\in\{1,2,\dots\},
\end{align*}
satisfies the following fractional equation which couples to the $(m-1)$ moment,
\begin{align*}
(-\Delta)^s v_m = mv_{m-1},
\end{align*}
with identical boundary conditions to the first moment and $v_1=v$. For the $m=2$ moment, this becomes
\begin{align}\label{m2}
(-\Delta)^s v_2 = 2v_{1}.
\end{align}
For $s\in(0,1/2]$, we have shown that $v_1(x)$ is constant in space to leading order, $v_1(x)\sim \mu_{s,\eps}$. Dividing \eqref{m2} by twice this constant implies that $w_2:=v_2/(2\mu_{s,\eps})$ satisfies the same fractional equation as the first moment $v_1$ to leading order. Hence, $w_2\sim v_1$ and thus $v_2\sim 2(v_1)^2$. Continuing this argument yields the leading order behavior of the $m$th moment,
\begin{align*}
    v_m\sim m!(v_1)^m,\quad m\in\{1,2,\dots\},
\end{align*}
which implies that $\tau/\mu_{s,\eps}$ is exponentially distributed with unit mean in the limit $\eps\to0$ (since exponential random variables are determined by their moments \cite{billingsley2008}).

\section{Numerical Simulations}\label{sec:numerical}

In this section we numerically calculate the FHT by solving the fractional differential equation \eqref{eq:hitting-time-pde} directly, as well as by using Monte-Carlo methods. These numerical calculations will serve the purpose of validating the formal asymptotic calculations of the previous section, with the Monte Carlo simulations also allowing us to investigate the full probability distribution of the FHT. We proceed by first outlining the numerical methods used to solve \eqref{eq:hitting-time-pde} in Section \ref{subsec:numerical-pde}. In Section \ref{subsec:monte-carlo} we outline the  methods used in the Monte-Carlo simulations. Finally, in Section \ref{subsec:numerical-results} we showcase the results from our numerical computations.

\subsection{Solving the MFHT Fractional Differential Equation}\label{subsec:numerical-pde}

To numerically solve \eqref{eq:hitting-time-pde} we require only a numerical discretization of the periodic fractional Laplacian $(-\Delta)^s$. Our numerical discretization of the periodic fractional Laplacian is based on the finite difference-quadrature approach of Huang and Oberman \cite{huang_2014}. Fix an integer $M>0$, let $h=2/M$, and let
\begin{equation}
    z_n = -1 + hn,\qquad n\in\mathscr{M}:=\{0,...,M-1\},
\end{equation}
be a uniform discretization of the interval $-1<x<1$. Denote by $(-\Delta_h)^s$ the numerical discretization of the periodic fractional Laplacian on $-1<x<1$. The discrete operator $(-\Delta_h)^s$ acts on an arbitrary vector $\pmb{\varphi} = (\varphi_0,...,\varphi_{M-1})^T$ according to (see equation ($\text{FL}_h$) in \cite{huang_2014})
\begin{equation}\label{eq:num-frac-lap}
    ((-\Delta_h)^s\pmb{\varphi})_n = \sum_{m\in\mathscr{M}}(\varphi_n-\varphi_m)W_{n-m},\quad W_{\sigma} := w_\sigma + \sum_{k=1}^\infty \left(w_{\sigma-kM}+w_{\sigma+kM}\right).
\end{equation}
where we have used used periodicity to simplify the expression, and where each $w_m$ ($m\in\mathbb{Z})$ is an appropriately chosen weight. See Appendix \ref{app:num-frac-lap} for additional details on our choice of weights, as well as some practical considerations for their computation. Define the set $\mathscr{I}:=\{n\in\mathscr{M}\,|\, z_n\in\cup_{i=1}^N(x_i-\ep l_i,x_i+\ep l_i)\}$. The numerical solution to the hitting-time problem \eqref{eq:hitting-time-pde} is then obtained by finding $\pmb{v} = (v_0,...,v_{M-1})^T$ satisfying linear system
\begin{equation}\label{eq:num-hitting}
    \begin{cases}
        \sum_{m\in\mathscr{M}\setminus\mathscr{I}} (v_n-v_m)W_{n-m} = 1, & n\in\mathscr{M}\setminus\mathscr{I},\\
        v_n = 0, & n\in\mathscr{I}.
    \end{cases}
\end{equation}
In Section \ref{subsec:numerical-results} we use $M=50,000$ points and $K=10,000$ terms in the evaluation of the weights $W_\sigma$ (see \eqref{eq:weight-fast-summation} in Appendix \ref{app:num-frac-lap}). 

\subsection{Monte Carlo}\label{subsec:monte-carlo}

We now describe the stochastic simulation algorithm used to generate FHTs of L{\'e}vy flights. Our stochastic simulation algorithm relies on constructing a L{\'e}vy fight by subordinating a Brownian motion \cite{lawley2023super}. Specifically, let $B=\{B(u)\}_{u\ge0}$ be a one-dimensional Brownian motion with unit diffusivity (i.e.\ scaled so that $\E[(B(u))^2]=2u$ for all $u\ge0$), and let $U=\{U(t)\}_{t\ge0}$ be an independent $s$-stable L{\'e}vy subordinator (i.e.\ it has Laplace exponent $\Phi(\beta)=\beta^s$). Then the following random time change of $B$,
\begin{align}\label{timechange}
    X(t)
    :=D_s^{1/(2s)}B(U(t))\quad t\ge0,
\end{align}
is a L\'evy flight with generalized diffusivity $D_s>0$.

Given a discrete time step $\Delta t>0$, we construct a statistically exact path of the $s$-stable subordinator $\{U(t)\}_{t\ge0}$ on the discrete time grid $\{t_{k}\}_{k\in\mathbb{N}}$ with $t_{k}=k\Delta t$ via
\begin{align*}
U(t_{k+1})
=U(t_{k})+(\Delta t)^{1/s}\Theta_{k},\quad k\ge0,
\end{align*}
where $U(t_{0})=U(0)=0$ and $\{\Theta_{k}\}_{k\in\mathbb{N}}$ is an iid sequence of realizations of \cite{carnaffan2017}
\begin{align*}
\Theta
=\frac{\sin(s(V+\pi/2)}{(\cos(V))^{1/s}}\bigg(\frac{\cos(V-\gamma(V+\pi/2))}{E}\bigg)^{(1-s)/s},
\end{align*}
where $V$ is uniformly distributed on $(-\pi/2,\pi/2)$ and $E$ is an independent exponential random variable with $\E[E]=1$. We then construct a statistically exact path of the Brownian motion $\{{{B}}(u)\}_{u\ge0}$ on the (random) discrete time grid $\{U(t_{k})\}_{k\in\mathbb{N}}$ via
\begin{align*}
{{B}}(U(t_{k+1}))
={{B}}(U(t_{k}))+\sqrt{2(\Delta t)^{1/s}\Theta_{k}}\xi_{k},\quad k\ge0,
\end{align*}
where $\{\xi_{k}\}_{k\in\mathbb{Z}}$ is an iid sequence of standard Gaussian random variables and we impose periodic boundary conditions. Finally, we obtain a statistically exact path of the L{\'e}vy flight ${{X}}=\{{{X}}(t)\}_{t\ge0}$ in \eqref{timechange} on the discrete time grid $\{t_{k}\}_{k\in\mathbb{N}}$ via ${{X}}(t_{k})
=D_s^{1/(2s)}{{B}}(U(t_{k}))$ for $k\ge0$. The FHT $\tau$ to the target set $U_{\textup{target}}$ is then approximated by $\tau\approx \overline{k}\Delta t$ where $\overline{k}
:=\min\{k\Delta t\ge0:{{X}}(t_{k})\in U_{\textup{target}}\}$.

The Monte Carlo data in the results below is computed from $10^{3}$ independent trials with $\Delta t=10^{-5}$ and $D_s=1$.

\subsection{Results}\label{subsec:numerical-results}

\begin{figure}[t!]
	\centering % <-- added
	\begin{subfigure}{0.33\textwidth}
		\includegraphics[width=\linewidth]{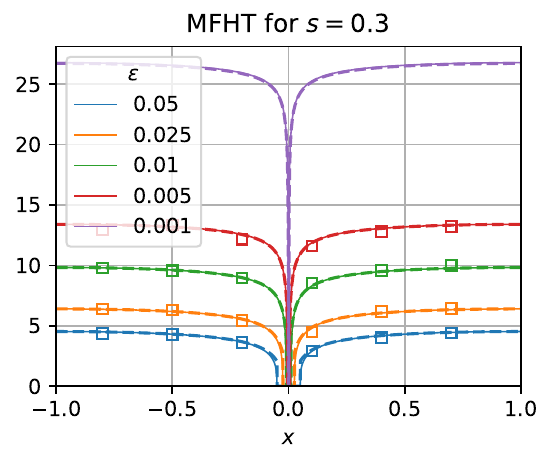}
	\end{subfigure}\hfil % <-- added
	\begin{subfigure}{0.33\textwidth}
		\includegraphics[width=\linewidth]{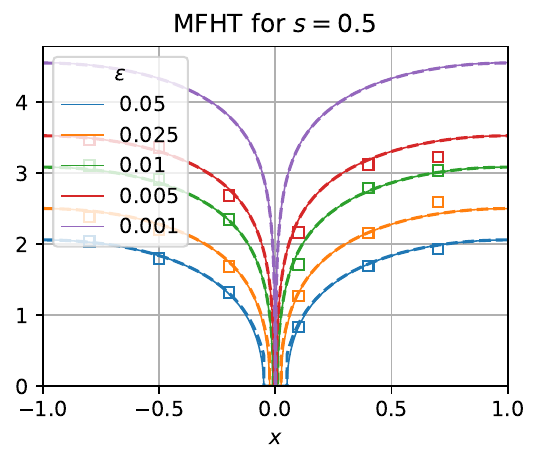}
	\end{subfigure}\hfil % <-- added
	\begin{subfigure}{0.33\textwidth}
		\includegraphics[width=\linewidth]{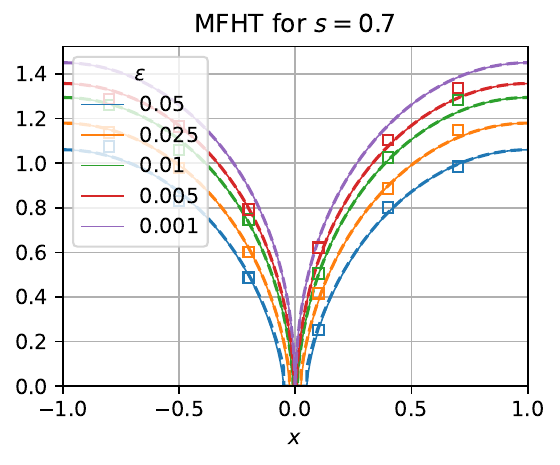}
	\end{subfigure}
	\caption{MFHT for the symmetric one-target configuration.  Solid curves, dashed curves, and hollow squares correspond to solutions obtained by numerically solving the fractional PDE \eqref{eq:hitting-time-pde}, by using the asymptotic approximations \eqref{eq:v-asy-sol}, and from Monte Carlo simulations respectively.}\label{fig:symmetric_hitting_time}
\end{figure}

To validate our asymptotic analysis we compare our asymptotic approximations for the MFHT with full numerical simulations using the methods outlined in Sections \ref{subsec:numerical-pde} and \ref{subsec:monte-carlo}. We present this comparison for two types of configurations. The first, which we refer to as the \textit{symmetric one-target configuration}, consists of a single target with $x_1=0$ and $l_1=1$. The second, which we refer to as the \textit{asymmetric three-target configuration}, consists of $N=3$ targets centred at $x_1=-0.6$, $x_2=0.4$, and $x_3=0.75$ with $l_1=1$, $l_2=1.25$, and $l_3=1.5$. 

\begin{figure}[t!]
	\centering % <-- added
	\begin{subfigure}{0.33\textwidth}
		\includegraphics[width=\linewidth]{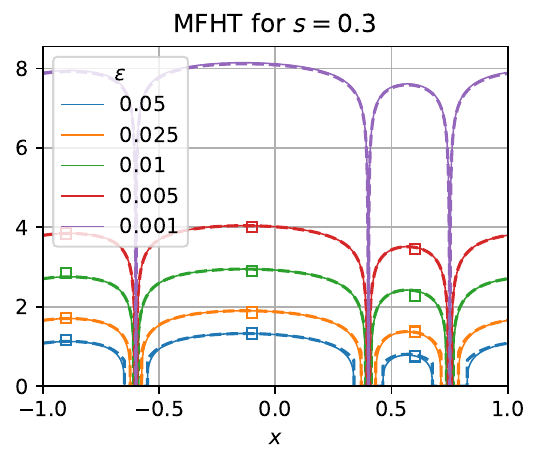}
	\end{subfigure}\hfil % <-- added
	\begin{subfigure}{0.33\textwidth}
		\includegraphics[width=\linewidth]{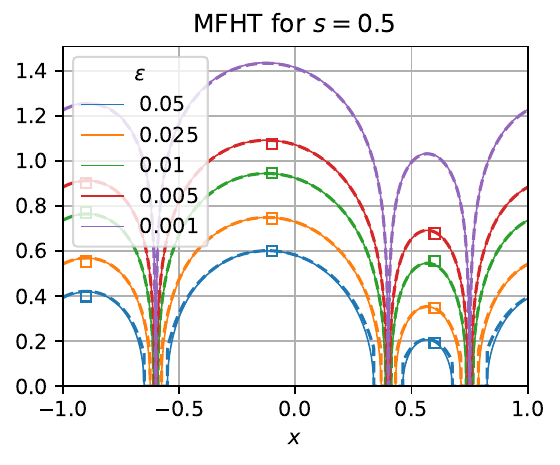}
	\end{subfigure}\hfil % <-- added
	\begin{subfigure}{0.33\textwidth}
		\includegraphics[width=\linewidth]{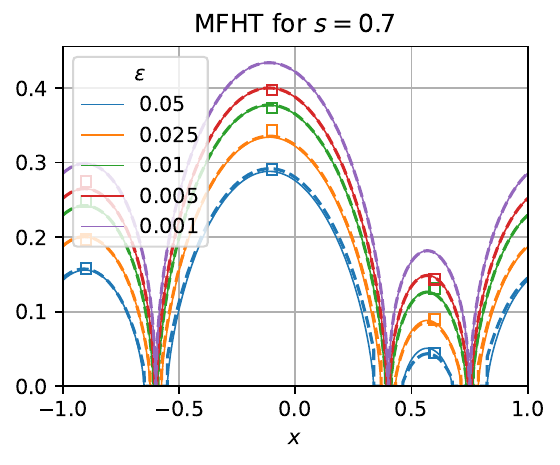}
	\end{subfigure}
	\caption{MFHT for the asymmetric three-target configuration. Solid curves, dashed curves, and hollow squares correspond to solutions obtained by numerically solving the fractional differential equation \eqref{eq:hitting-time-pde}, by using the asymptotic approximations \eqref{eq:v-asy-sol}, and from Monte Carlo simulations respectively.}\label{fig:asymmetric_hitting_time}
\end{figure}

In Figures \ref{fig:symmetric_hitting_time} and \ref{fig:asymmetric_hitting_time} we plot the MFHT for the symmetric one-target and asymmetric three-target configurations respectively. Specifically, each figure compares the solution obtained by solving \eqref{eq:hitting-time-pde} numerically (solid curves), the solution obtained using the asymptotic approximation \eqref{eq:v-asy-sol} (dashed curves), as well as the values of the MFHT starting from specific values of $x\in(-1,1)$ obtained from Monte Carlo simulations (hollow squares). In each case we observe excellent agreement between the asymptotic and numerical solutions even for moderately sized values of $\varepsilon>0$. In addition to validating our asymptotic approximations, the plots in Figures \ref{fig:symmetric_hitting_time} and \ref{fig:asymmetric_hitting_time} also showcase the qualitative properties of the MFHT predicted by our asymptotic analysis. Specifically, they illustrate a strong $\varepsilon$-dependence when $s<1/2$ in contrast to when $s>1/2$ which supports the scaling $v=O(\varepsilon^{2s-1})$ for $s<1/2$ and $v=O(1)$ for $s>1/2$. Moreover, we observe that for sufficiently small values of $\varepsilon>0$, the MFHT in the outer region is approximately spatially constant when $s<1/2$ whereas it is spatially variable when $s>1/2$. Although the leading order asymptotics predict a spatially constant solution for $s=1/2$, this is difficult to see numerically since the first order correction is $O(1/\log\ep)$.

\begin{figure}[t!]
	\centering % <-- added
	\begin{subfigure}{0.5\textwidth}
		\includegraphics[width=\linewidth]{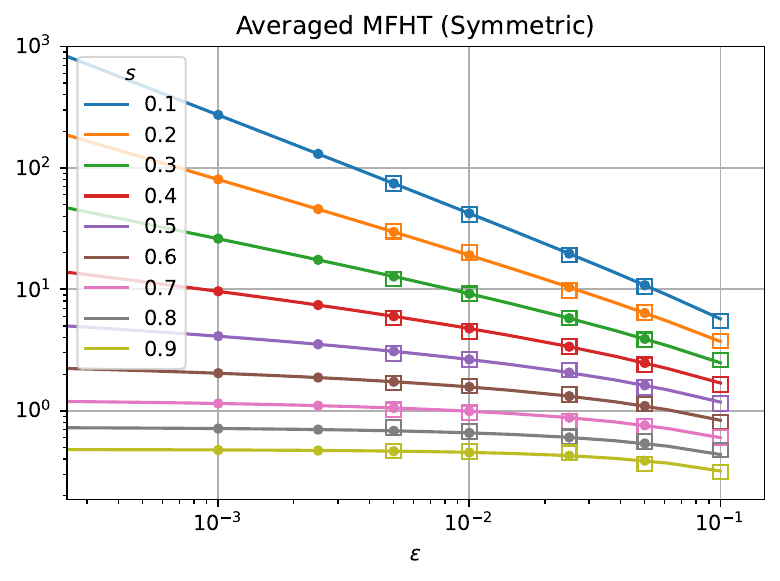}
	\end{subfigure}\hfil % <-- added
	\begin{subfigure}{0.5\textwidth}
		\includegraphics[width=\linewidth]{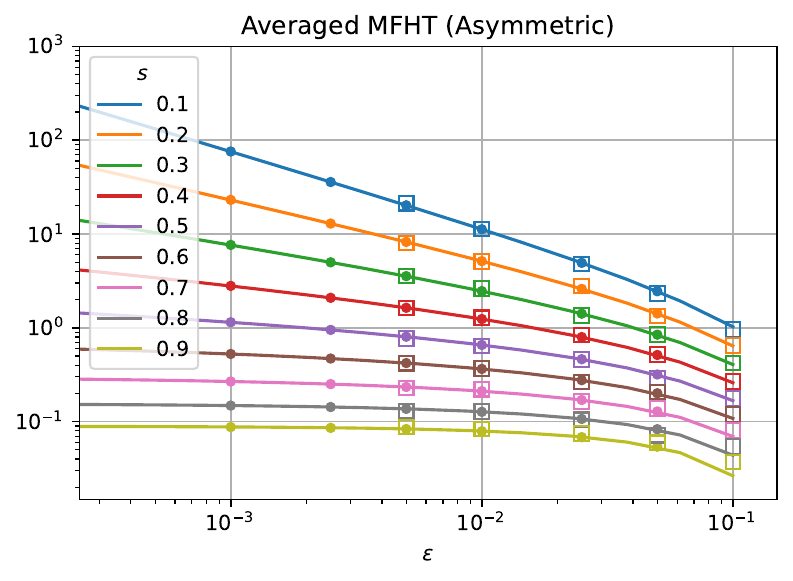}
	\end{subfigure}
	\caption{The MFHT averaged over a uniformly distributed initial condition in $\Omega\setminus\Omega_\text{target}$ for the (left) symmetric one-target configuration, and the (right) asymmetric three-target configuration. In each plot the solid curve indicates the asymptotic approximation, the dots indicate results from numerically solving the fractional differential equation \ref{eq:hitting-time-pde}, and the hollow squares those values obtained by stochastic simulations.}\label{fig:ave_hitting_time}
\end{figure}

An additional quantity of interest is the MFHT averaged over uniformly distributed initial points $x\in(-1,1)$, i.e.
\begin{equation*}
    \overline{v} := \frac{1}{2}\int_{-1}^{1}v(x)dx.
\end{equation*}
In Figure \ref{fig:ave_hitting_time} we plot this averaged MFHT versus $\varepsilon>0$ for different values of $0<s<1$ for both the symmetric one-target and asymmetric three-target configurations. In each plot the solid curve corresponds to the asymptotically computed solution which, in light of the vanishing integral constraint in \eqref{eq:greens-pde}, is equal to $\chi^\varepsilon$ given by \eqref{eq:chi-def}. The solid dots correspond to values obtained by numerically integrating the numerical solution to \eqref{eq:hitting-time-pde}, whereas the hollow squares are results from Monte Carlo simulations. These plots shows good agreement between the asymptotic approximation and numerical simulations.

\begin{figure}[t!]
	\centering % <-- added
	\begin{subfigure}{0.33\textwidth}
		\includegraphics[width=\linewidth]{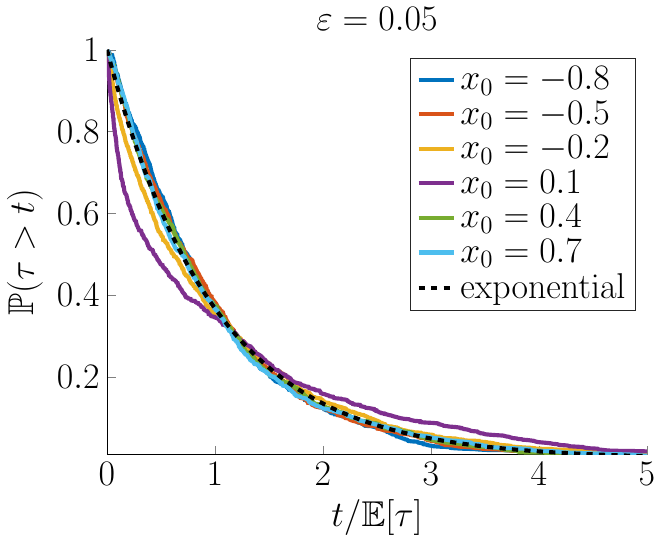}
		%\caption{$\eps=0.05$}\label{figdist1}
	\end{subfigure}\hfil % <-- added
	\begin{subfigure}{0.33\textwidth}
		\includegraphics[width=\linewidth]{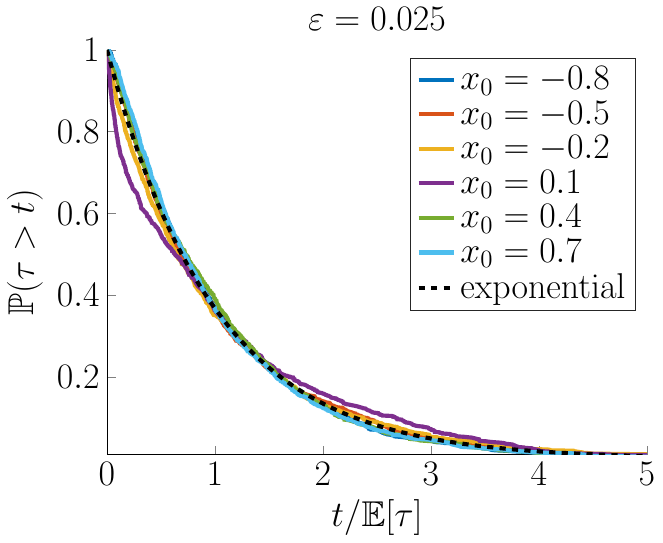}
		%\caption{$\eps=0.025$}\label{figdist2}
	\end{subfigure}\hfil % <-- added
	\begin{subfigure}{0.33\textwidth}
		\includegraphics[width=\linewidth]{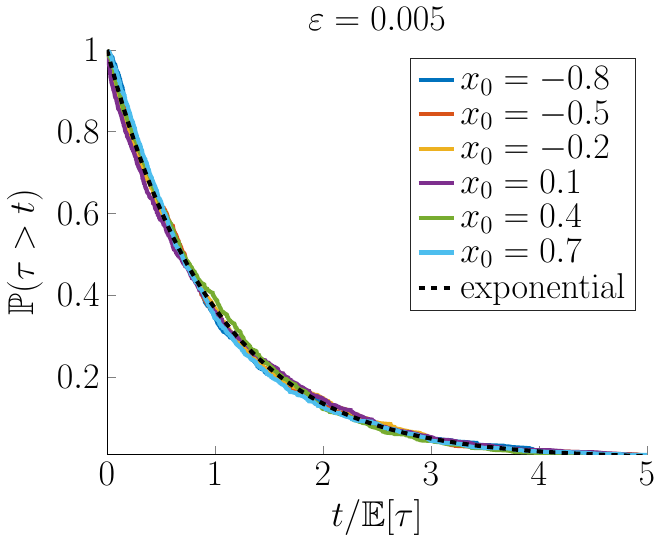}
		%\caption{$\eps=0.005$}\label{figdist3}
	\end{subfigure}
	\caption{
 Probability distribution of FHT.
 }\label{figdist}
\end{figure}

Finally, in Figure~\ref{figdist}, we compare (i) the full probability distribution of the FHT $\tau$ computed from stochastic simulations to (ii) the exponential distribution implied by the analysis in section~\ref{subsec:higher-moments}. This plot is for the symmetric one-target configuration in Figure~\ref{fig:symmetric_hitting_time} with $s=0.3$. The convergence to an exponential distribution is apparent as $\eps$ decreases from $\eps=0.05$ in the left panel down to $\eps=0.005$ in the right panel.

\section{Optimal random search}\label{sectionopt}

We now investigate the value of the fractional order $s\in(0,1]$ which minimizes the averaged MFHT. By averaging over a uniformly distributed initial position, considering the case $N=1$, neglecting the highest order terms from our asymptotic expansion, and  reversing the nondimensionalization in \eqref{rescale}, we arrive at the following dimensional measure of the search time,
\begin{align*}
T_s
:=\displaystyle\begin{dcases}
(l^{2s}/D_s)(\eps^{2s-1}2\mathfrak{a}_{s}/\mathfrak{b}_{s}
-2\mathfrak{a}_{s}R_{s}(0)) & \text{if }s\neq1/2,\\
(l^{2s}/D_s)(\log(2/\eps)2/\pi
-2R_{1/2}(0)/\pi)  & \text{if }s=1/2,
\end{dcases}\qquad \textup{for }s\in(0,1).
\end{align*}
That is, $T_s$ is the averaged MFHT over uniformly distributed initial positions of a one-dimensional, $(2s)$-stable L{\'e}vy flight with generalized diffusivity $D_s>0$ 
, and an infinite periodic array of targets with separation distance $2l>0$ where each target has radius $\eps l$ with $0<\eps\ll1$.

To study how $T_s$ depends on $s\in(0,1]$, we must choose how the generalized diffusivity $D_s$ depends on $s$ (since it has dimension $[D_s]=(\textup{length})^{2s}/(\textup{time})$). We follow \cite{palyulin2014} and introduce a lengthscale $l_0>0$ (independent of $s$) and suppose
\begin{align*}
D_s=(l_0)^{2s}/t_0    
\end{align*}
for some timescale $t_0$. Such a lengthscale $l_0>0$ arises naturally in the continuous-time random walk derivation of a L{\'e}vy flight (see \eqref{pl}-\eqref{Ds} in Section \ref{sec:intro} and \cite{metzler2004} for more details). Normalizing $T_s$ by the Brownian search time $T_1:=(l^{2}/D_1)(1-\eps)^2/3$ yields the following ratio for $s\in(0,1)$,
\begin{align}\label{rho}
    \rho(s)
    :=\frac{T_s}{T_1}
    =\frac{(l_0/l)^{2(1-s)}}{(1-\eps)^2/3}\times\displaystyle\begin{dcases}
\eps^{2s-1}2\mathfrak{a}_{s}/\mathfrak{b}_{s}
-2\mathfrak{a}_{s} R_{s}(0)& \text{if }s\neq1/2,\\
\log(2/\eps)2/\pi
-2R_{1/2}(0)/\pi  & \text{if }s=1/2.
\end{dcases}
\end{align}
Hence, $\rho(s)<1$ (respectively $\rho(s)>1$) means that the L{\'e}vy search is faster (respectively slower) than Brownian search.

\begin{figure}[t!]
	\centering % <-- added
	\begin{subfigure}{0.49\textwidth}
		\includegraphics[width=\linewidth]{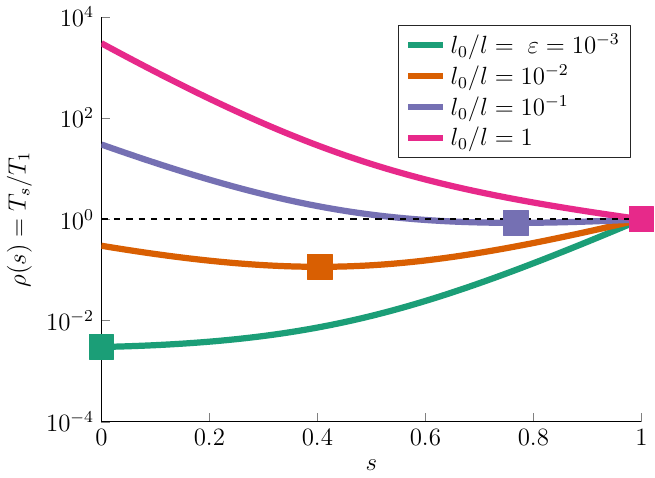}
  \label{figsa}
	\end{subfigure}\hfil % <-- added
	\begin{subfigure}{0.49\textwidth}
		\includegraphics[width=\linewidth]{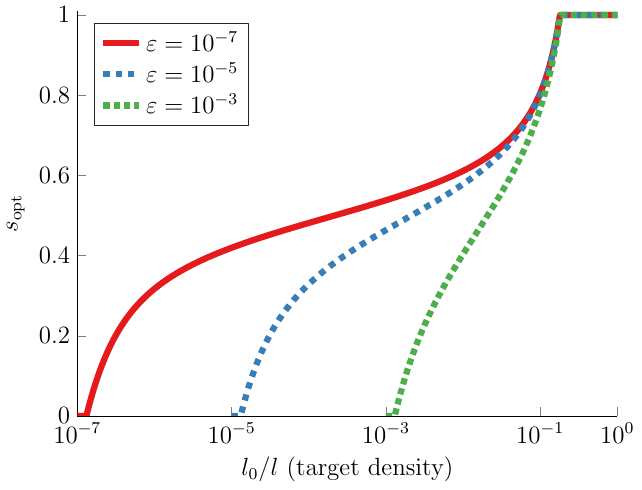}
	\end{subfigure}\hfil % <-- added
	\caption{
(left) $\rho(s)$ in \eqref{rho} as a function of $s\in(0,1)$ for $\eps=10^{-3}$ and different values of the target density $l_0/l$. Square markers indicate $s_{\textup{opt}}$ in \eqref{sopt}. (right) $s_{\textup{opt}}$ as a function of the target density $l_0/l$ for different values of $\eps$.
 }\label{figs}
\end{figure}

In the left panel of Figure~\ref{figs}, we plot $\rho(s)$ as a function of $s\in(0,1)$ for different values of $l_0/l$. Notice that $l_0/l\ll1$ describes sparse targets and $l_0/l\not\ll1$ describes dense targets (where ``sparse'' and ``dense'' are relative to the lengthscale $l_0$). This plot shows that L{\'e}vy search is faster than Brownian search for sparse targets, whereas Brownian search is faster than L{\'e}vy search for dense targets.

In the right panel of Figure~\ref{figs}, we plot the ``optimal'' value of $s\in(0,1]$ which minimizes the search time,
\begin{align}\label{sopt}
    s_{\textup{opt}}
    :=\underset{s}{\rm arg\,min}\;\rho(s),
\end{align}
as a function of the target density $l_0/l$ for fixed values of $\eps$. This plot shows that $s_{\textup{opt}}$ varies continuously from $s_{\textup{opt}}\approx0$ for sparse targets up to $s_{\textup{opt}}\approx1$ (i.e.\ Brownian search) as the target density increases. Hence, the value $s=1/2$ (which corresponds to stability index $\alpha=2s=1$, i.e.\ so-called inverse square L\'evy search) is not distinguished from other values of $s\in(0,1]$ in the sense that $s_{\textup{opt}}=1/2$ for only a single value of the target density $l_0/l$ for each $\eps>0$. On the other hand, we do find that $s_{\text{opt}}\to1/2$ if we first take the limit $\eps\to0$ and then take $l_0/l\to0$. To see this, note first that we must have $\lim_{\ep\rightarrow0}s_{\textup{opt}}>1/2$ since \eqref{rho} implies $\lim_{\eps\to0}\rho(s)=+\infty$ if $s\le1/2$. Next, \eqref{rho} implies 
\begin{align*}
    \lim_{\eps\to0}\rho(s)=-((l_0/l)^{2(1-s)}/3)(2\mathfrak{a}_sR_s(0))>0\quad\text{if }s>1/2,
\end{align*}
and therefore $\lim_{l_0/l\to0}\lim_{\eps\to0}s_{\textup{opt}}=1/2$.

% (Figure~\ref{figs} shows $s_{\text{opt}}\approx1/2$ if $l_0/l\approx\sqrt{\eps}\ll1$). 
%However, the rand of target densities $l_0/l$ for which $s_{\textup{opt}}\approx1/2$ grows as $\eps$ decays. For instance, Figure~\ref{figs} shows that $s_{\textup{opt}}\in($

%Furthermore, Figure~\ref{figs} suggests that
%\begin{align*}
%    s_{\text{opt}}\approx1/2\quad\text{if }\eps\approx(l_0/l)^2\ll1.
%\end{align*}
%To see
%To summarize, if $\eps\ll1$, then
%\begin{alignat*}{2}
%    s_{\textup{opt}}&\approx1\quad&&\text{if }l_0/l\not\ll1,\\
 %   s_{\textup{opt}}&\approx1/2\quad&&\text{if }\eps\ll l_0/l\ll1,\\
  %  s_{\textup{opt}}&\approx0\quad&&\text{if }\eps\not\ll l_0/l\ll1,\\
%\end{alignat*}

%To estimate the rate of convergence, note that Proposition~\ref{prop:greens-func} implies $\lim_{s\to1/2}R_s(0)\to1$ and \eqref{eq:def_of_a_and_b} implies $\mathfrak{a}_s\sim-1/(2\pi(s-1/2))$ as $s\to1/2+$, and thus it follows that 
%\begin{align*}
    %s_{\textup{opt}}^{(0)}
    %\sim\frac{1}{2}(1-1/\log(l_0/l))\quad\text{as }l_0/l\to0,
%\end{align*}
%which indicates that the convergence in \eqref{0opt} is quite slow. [[Seem to need
%\begin{align*}
    %\Big(\frac{\eps}{l_0/l}\Big)^{-1/\log(l_0/l)}\ll1\quad\text{and}\quad -1/\log(l_0/l)\ll1,
%\end{align*}
%which is a pretty small region of parameter space.]]

\section{Discussion}

In this paper we calculated an asymptotic approximation for the MFHT to a small target in a periodic one-dimensional domain. Our asymptotic approximation is summarized in Principal Result 1 and reduces the calculation of the MFHT to that of solving the linear system \eqref{eq:B-sys}, thereby providing a fast method for approximating the MFHT when the target size is small. In the special case of a symmetric configuration it suffices to consider the case a single target for which the system \eqref{eq:B-sys} can be solved explicitly (see \eqref{s1}--\eqref{unif0} in Section \ref{sec:intro}). Furthermore we validated our asymptotics by comparing them to numerical computations of the MFHT obtained by solving the fractional differential equation \eqref{eq:hitting-time-pde} directly and by using stochastic simulations. 

The asymptotic analysis leading to Principal Result 1 is analogous to that used in two- and three-dimensional narrow capture/escape problems involving pure diffusion \cite{ward10,ward10b}. This analogy was previously identified in \cite{gomez_medeiros_2022,gomez_2023} and is a result of the singular behaviour of the fractional free-space Green's function which is logarithmic when $s=1/2$ and algebraic when $s<1/2$, mirroring that of the classical free-space Green's function in two- and three-dimensions respectively. A novel aspect of the asymptotic analysis presented in this paper is the recognition of a fractional counterpart to the classical electrified disk problem. This fractional differential equation was solved by using a fractional Kelvin transform and fractional Poisson kernel for $s\neq 1/2$, and by considering a two-dimensional extended problem solvable by complex analysis methods for $s=1/2$. In addition, we determined that when $s\leq 1/2$ the MFHT is spatially constant to leading order, with this observation further allowing us to conclude that the FHT is exponentially distributed when $s\leq 1/2$.

The present study joins many prior works which use L\'evy flights as simple theoretical models to investigate optimal search strategies. 
%There is a long line of previous works investigating L\'evy search processes. 
%L\'evy flights are often used as simple theoretical models to investigate optimal search strategies. 
Prior works often choose one-dimensional spatial domains due to their analytical tractability and as models for search in effectively one-dimensional domains such as streams, along coastlines, at forest-meadows, and other borders \cite{palyulin2014, kusmierz2015, palyulin2016, palyulin2017, palyulin2019, padash2020, padash2022}. The very interesting work of Palyulin, Chechkin, and Metzler \cite{palyulin2014} is perhaps most closely related to our present study. In \cite{palyulin2014}, the authors consider a one-dimensional, possibly biased L\'evy flight on the entire real line with a single point-like target. A major result of \cite{palyulin2014} is that despite the frequent claim that L\'evy flights with $s=1/2$ are most efficient for sparse targets, the optimal value of $s$ may range the entire interval between $s=1/2$ and $s=1$ and thus include Brownian search (the assumption of a point-like target in \cite{palyulin2014} meant that these authors did not consider $s<1/2$). Indeed, as the authors of \cite{palyulin2014} state, ``the main message from this study is that L\'evy flight search and its optimization is sensitive to the exact conditions'' and ``our results show clear limitations for the universality of L\'evy flight foraging'' \cite{palyulin2014}. Our results agree with these main points, as the optimal value of $s$ in our study spans the entire interval $(0,1]$ as the target density $l_0/l$ increases from $l_0/l\le\eps$ up to $l_0/l\approx1$ (see Figure~\ref{figs}).

\bibliographystyle{siam}
\bibliography{biblio}

\begin{thebibliography}{10}

\bibitem{abatangelo_2019}
{\sc N.~Abatangelo and E.~Valdinoci}, {\em Getting acquainted with the
  fractional laplacian}, Contemporary research in elliptic PDEs and related
  topics,  (2019), pp.~1--105.

\bibitem{Ammari2011}
{\sc H.~Ammari, J.~Garnier, H.~Kang, H.~Lee, and K.~S{\o}lna}, {\em The mean
  escape time for a narrow escape problem with multiple switching gates},
  Multiscale Model Simul, 9 (2011), pp.~817--833.

\bibitem{benichou2010}
{\sc O.~Benichou, D.~Grebenkov, P.~Levitz, C.~Loverdo, and R.~Voituriez}, {\em
  Optimal {Reaction} {Time} for {Surface}-{Mediated} {Diffusion}}, Phys Rev
  Lett, 105 (2010), p.~150606.

\bibitem{benichou2011rev}
{\sc O.~B{\'e}nichou, C.~Loverdo, M.~Moreau, and R.~Voituriez}, {\em
  Intermittent search strategies}, Rev Mod Phys, 83 (2011), p.~81.

\bibitem{bertoin1996}
{\sc J.~Bertoin}, {\em {L}{\'e}vy processes}, vol.~121, Cambridge {U}niversity
  {P}ress, 1996.

\bibitem{billingsley2008}
{\sc P.~Billingsley}, {\em Probability and measure}, John Wiley \& Sons, 2008.

\bibitem{blumenthal_1961}
{\sc R.~M. Blumenthal, R.~K. Getoor, and D.~Ray}, {\em On the distribution of
  first hits for the symmetric stable processes.}, Transactions of the American
  Mathematical Society, 99 (1961), pp.~540--554.

\bibitem{bressloff2017mean}
{\sc P.~C. Bressloff and S.~D. Lawley}, {\em Mean first passage times for
  piecewise deterministic markov processes and the effects of critical points},
  Journal of Statistical Mechanics: Theory and Experiment, 2017 (2017),
  p.~063202.

\bibitem{bressloff2022}
{\sc P.~C. Bressloff and R.~D. Schumm}, {\em The narrow capture problem with
  partially absorbing targets and stochastic resetting}, Multiscale Modeling \&
  Simulation, 20 (2022), pp.~857--881.

\bibitem{bucur_2016}
{\sc C.~Bucur}, {\em Some observations on the {G}reen function for the ball in
  the fractional {L}aplace framework}, Communications on Pure and Applied
  Analysis, 15 (2016), pp.~657--699.

\bibitem{buldyrev2021}
{\sc S.~Buldyrev, E.~Raposo, F.~Bartumeus, S.~Havlin, F.~Rusch, M.~da~Luz, and
  G.~Viswanathan}, {\em {C}omment on ``{I}nverse square {L}{\'e}vy walks are
  not optimal search strategies for $d\ge2$''}, Physical Review Letters, 126
  (2021), p.~048901.

\bibitem{caffarelli_2007}
{\sc L.~Caffarelli and L.~Silvestre}, {\em An extension problem related to the
  fractional laplacian}, Communications in partial differential equations, 32
  (2007), pp.~1245--1260.

\bibitem{carnaffan2017}
{\sc S.~Carnaffan and R.~Kawai}, {\em Solving multidimensional fractional
  {F}okker--{P}lanck equations via unbiased density formulas for anomalous
  diffusion processes}, SIAM Journal on Scientific Computing, 39 (2017),
  pp.~B886--B915.

\bibitem{chaubet2022}
{\sc Y.~Chaubet, T.~Lefeuvre, Y.~G. Bonthonneau, and L.~Tzou}, {\em {Geodesic
  Levy Flights and Expected Stopping Time for Random Searches}}, arXiv preprint
  arXiv:2211.13973,  (2022).

\bibitem{chechkin2003}
{\sc A.~V. Chechkin, R.~Metzler, V.~Y. Gonchar, J.~Klafter, and L.~V.
  Tanatarov}, {\em First passage and arrival time densities for l{\'e}vy
  flights and the failure of the method of images}, Journal of Physics A:
  Mathematical and General, 36 (2003), p.~L537.

\bibitem{ward10b}
{\sc A.~F. Cheviakov, M.~J. Ward, and R.~Straube}, {\em An asymptotic analysis
  of the mean first passage time for narrow escape problems: {P}art {II:} {T}he
  sphere}, Multiscale Model Simul., 8 (2010), pp.~836--870.

\bibitem{chou_first_2014}
{\sc T.~Chou and M.~R. D'Orsogna}, {\em First passage problems in biology}, in
  First-{Passage} {Phenomena} and {Their} {Applications}, World Scientific,
  2014, pp.~306--345.

\bibitem{NIST:DLMF}
{\em {\it NIST Digital Library of Mathematical Functions}}.
\newblock http://dlmf.nist.gov/, Release 1.0.28 of 2020-09-15.
\newblock F.~W.~J. Olver, A.~B. {Olde Daalhuis}, D.~W. Lozier, B.~I. Schneider,
  R.~F. Boisvert, C.~W. Clark, B.~R. Miller, B.~V. Saunders, H.~S. Cohl, and
  M.~A. McClain, eds.

\bibitem{dubkov2008}
{\sc A.~A. Dubkov, B.~Spagnolo, and V.~V. Uchaikin}, {\em {L}{\'e}vy flight
  superdiffusion: an introduction}, International Journal of Bifurcation and
  Chaos, 18 (2008), pp.~2649--2672.

\bibitem{durrett2019}
{\sc R.~Durrett}, {\em Probability: theory and examples}, Cambridge university
  press, 2019.

\bibitem{gomez_medeiros_2022}
{\sc D.~Gomez, M.~De~Medeiros, J.-c. Wei, and W.~Yang}, {\em Spike solutions to
  the supercritical fractional gierer-meinhardt system}, 2023.

\bibitem{gomez_2023}
{\sc D.~Gomez, J.~Wei, and Z.~Yang}, {\em Multi-spike solutions to the
  one-dimensional subcritical fractional {S}chnakenberg system}, Physica D:
  Nonlinear Phenomena, 448 (2023), p.~133720.

\bibitem{grebenkov2017narrow}
{\sc D.~S. Grebenkov and G.~Oshanin}, {\em Diffusive escape through a narrow
  opening: new insights into a classic problem}, Phys Chem Chem Phys, 19
  (2017), pp.~2723--2739.

\bibitem{guinard2021}
{\sc B.~Guinard and A.~Korman}, {\em Intermittent inverse-square l{\'e}vy walks
  are optimal for finding targets of all sizes}, Science advances, 7 (2021),
  p.~eabe8211.

\bibitem{holcman2014}
{\sc D.~Holcman and Z.~Schuss}, {\em The narrow escape problem}, {SIAM} Rev, 56
  (2014), pp.~213--257.

\bibitem{huang_2014}
{\sc Y.~Huang and A.~Oberman}, {\em Numerical methods for the fractional
  laplacian: A finite difference-quadrature approach}, SIAM Journal on
  Numerical Analysis, 52 (2014), pp.~3056--3084.

\bibitem{jolakoski2022}
{\sc P.~Jolakoski, A.~Pal, T.~Sandev, L.~Kocarev, R.~Metzler, and
  V.~Stojkoski}, {\em The fate of the american dream: A first passage under
  resetting approach to income dynamics}, arXiv preprint arXiv:2212.13176,
  (2022).

\bibitem{kaye2020}
{\sc J.~Kaye and L.~Greengard}, {\em A fast solver for the narrow capture and
  narrow escape problems in the sphere}, Journal of Computational Physics: X, 5
  (2020), p.~100047.

\bibitem{koren2007first}
{\sc T.~Koren, A.~Chechkin, and J.~Klafter}, {\em On the first passage time and
  leapover properties of {L}{\'e}vy motions}, Physica A: Statistical Mechanics
  and its Applications, 379 (2007), pp.~10--22.

\bibitem{koren2007}
{\sc T.~Koren, M.~A. Lomholt, A.~V. Chechkin, J.~Klafter, and R.~Metzler}, {\em
  Leapover lengths and first passage time statistics for {L}{\'e}vy flights},
  Physical review letters, 99 (2007), p.~160602.

\bibitem{kurella2015}
{\sc V.~Kurella, J.~C. Tzou, D.~Coombs, and M.~J. Ward}, {\em Asymptotic
  analysis of first passage time problems inspired by ecology}, Bull Math Biol,
  77 (2015), pp.~83--125.

\bibitem{kusmierz2015}
{\sc {\L}.~Ku{\'s}mierz and E.~Gudowska-Nowak}, {\em Optimal first-arrival
  times in l{\'e}vy flights with resetting}, Physical Review E, 92 (2015),
  p.~052127.

\bibitem{lawley2023super}
{\sc S.~D. Lawley}, {\em {Extreme statistics of superdiffusive L{\'e}vy flights
  and every other L{\'e}vy subordinate Brownian motion}}, Journal of Nonlinear
  Science, 33 (2023), p.~53.

\bibitem{lawley2023bor}
{\sc S.~D. Lawley and J.~Johnson}, {\em Why is there an “oversupply” of
  human ovarian follicles?}, Biology of Reproduction, 108 (2023), pp.~814--821.

\bibitem{levernier2020}
{\sc N.~Levernier, J.~Textor, O.~B{\'e}nichou, and R.~Voituriez}, {\em Inverse
  square {L}{\'e}vy walks are not optimal search strategies for $d\ge 2$},
  Physical review letters, 124 (2020), p.~080601.

\bibitem{levernier2021reply}
\leavevmode\vrule height 2pt depth -1.6pt width 23pt, {\em {R}eply to
  ``{C}omment on `{I}nverse square {L}{\'e}vy walks are not optimal search
  strategies for $d\ge2$'''}, Physical Review Letters, 126 (2021), p.~048902.

\bibitem{lindsay2017}
{\sc A.~E. Lindsay, A.~J. Bernoff, and M.~J. Ward}, {\em First passage
  statistics for the capture of a brownian particle by a structured spherical
  target with multiple surface traps}, Multiscale Model Simul, 15 (2017),
  pp.~74--109.

\bibitem{lischke2020}
{\sc A.~Lischke, G.~Pang, M.~Gulian, F.~Song, C.~Glusa, X.~Zheng, Z.~Mao,
  W.~Cai, M.~M. Meerschaert, M.~Ainsworth, et~al.}, {\em What is the fractional
  {L}aplacian? {A} comparative review with new results}, Journal of
  Computational Physics, 404 (2020), p.~109009.

\bibitem{lomholt2005}
{\sc M.~A. Lomholt, T.~Ambj{\"o}rnsson, and R.~Metzler}, {\em Optimal target
  search on a fast-folding polymer chain with volume exchange}, Physical review
  letters, 95 (2005), p.~260603.

\bibitem{lomholt2008}
{\sc M.~A. Lomholt, K.~Tal, R.~Metzler, and K.~Joseph}, {\em L{\'e}vy
  strategies in intermittent search processes are advantageous}, Proceedings of
  the National Academy of Sciences, 105 (2008), pp.~11055--11059.

\bibitem{meerschaert2011}
{\sc M.~M. Meerschaert and A.~Sikorskii}, {\em Stochastic models for fractional
  calculus}, vol.~43, Walter de Gruyter GmbH \& Co KG, 2011.

\bibitem{metzler2004}
{\sc R.~Metzler and J.~Klafter}, {\em The restaurant at the end of the random
  walk: recent developments in the description of anomalous transport by
  fractional dynamics}, Journal of Physics A: Mathematical and General, 37
  (2004), p.~R161.

\bibitem{mirny2009}
{\sc L.~Mirny, M.~Slutsky, Z.~Wunderlich, A.~Tafvizi, J.~Leith, and
  A.~Kosmrlj}, {\em How a protein searches for its site on dna: the mechanism
  of facilitated diffusion}, Journal of Physics A: Mathematical and
  Theoretical, 42 (2009), p.~434013.

\bibitem{montroll1965}
{\sc E.~W. Montroll and G.~H. Weiss}, {\em Random walks on lattices. ii},
  Journal of Mathematical Physics, 6 (1965), pp.~167--181.

\bibitem{padash2020}
{\sc A.~Padash, A.~V. Chechkin, B.~Dybiec, M.~Magdziarz, B.~Shokri, and
  R.~Metzler}, {\em First passage time moments of asymmetric l{\'e}vy flights},
  Journal of Physics A: Mathematical and Theoretical, 53 (2020), p.~275002.

\bibitem{padash2019}
{\sc A.~Padash, A.~V. Chechkin, B.~Dybiec, I.~Pavlyukevich, B.~Shokri, and
  R.~Metzler}, {\em First-passage properties of asymmetric l{\'e}vy flights},
  Journal of Physics A: Mathematical and Theoretical, 52 (2019), p.~454004.

\bibitem{padash2022}
{\sc A.~Padash, T.~Sandev, H.~Kantz, R.~Metzler, and A.~V. Chechkin}, {\em
  Asymmetric l{\'e}vy flights are more efficient in random search}, Fractal and
  Fractional, 6 (2022), p.~260.

\bibitem{palyulin2019}
{\sc V.~V. Palyulin, G.~Blackburn, M.~A. Lomholt, N.~W. Watkins, R.~Metzler,
  R.~Klages, and A.~V. Chechkin}, {\em First passage and first hitting times of
  {L}{\'e}vy flights and {L}{\'e}vy walks}, New Journal of Physics, 21 (2019),
  p.~103028.

\bibitem{palyulin2016}
{\sc V.~V. Palyulin, A.~V. Chechkin, R.~Klages, and R.~Metzler}, {\em Search
  reliability and search efficiency of combined l{\'e}vy--brownian motion: long
  relocations mingled with thorough local exploration}, Journal of Physics A:
  Mathematical and Theoretical, 49 (2016), p.~394002.

\bibitem{palyulin2014}
{\sc V.~V. Palyulin, A.~V. Chechkin, and R.~Metzler}, {\em L{\'e}vy flights do
  not always optimize random blind search for sparse targets}, Proceedings of
  the National Academy of Sciences, 111 (2014), pp.~2931--2936.

\bibitem{palyulin2017}
{\sc V.~V. Palyulin, V.~N. Mantsevich, R.~Klages, R.~Metzler, and A.~V.
  Chechkin}, {\em Comparison of pure and combined search strategies for single
  and multiple targets}, The European Physical Journal B, 90 (2017), pp.~1--16.

\bibitem{pavlyukevich2007}
{\sc I.~Pavlyukevich}, {\em {L}{\'e}vy flights, non-local search and simulated
  annealing}, Journal of Computational Physics, 226 (2007), pp.~1830--1844.

\bibitem{pavlyukevich2008}
\leavevmode\vrule height 2pt depth -1.6pt width 23pt, {\em Simulated annealing
  for {L}{\'e}vy-driven jump-diffusions}, Stochastic processes and their
  applications, 118 (2008), pp.~1071--1105.

\bibitem{ward10}
{\sc S.~Pillay, M.~J. Ward, A.~Peirce, and T.~Kolokolnikov}, {\em {An
  asymptotic analysis of the mean first passage time for narrow escape
  problems: Part I: Two-dimensional domains}}, Multiscale Model Simul., 8
  (2010), pp.~803--835.

\bibitem{ralf2014}
{\sc M.~Ralf, R.~Sidney, and O.~Gleb}, {\em First-passage phenomena and their
  applications}, vol.~35, World Scientific, 2014.

\bibitem{redner2001}
{\sc S.~Redner}, {\em A guide to first-passage processes}, Cambridge University
  Press, 2001.

\bibitem{reva2021}
{\sc M.~Reva, D.~A. DiGregorio, and D.~S. Grebenkov}, {\em A first-passage
  approach to diffusion-influenced reversible binding and its insights into
  nanoscale signaling at the presynapse}, Scientific reports, 11 (2021),
  pp.~1--17.

\bibitem{oton_2014}
{\sc X.~Ros-Oton and J.~Serra}, {\em The dirichlet problem for the fractional
  laplacian: regularity up to the boundary}, Journal de Math{\'e}matiques Pures
  et Appliqu{\'e}es, 101 (2014), pp.~275--302.

\bibitem{tzou2023}
{\sc J.~Tzou and L.~Tzou}, {\em {Challenging the Levy Flight Foraging
  Hypothesis-A Joint Monte Carlo and Numerical PDE Approach}}, arXiv preprint
  arXiv:2302.13976,  (2023).

\bibitem{vaccario2015}
{\sc G.~Vaccario, C.~Antoine, and J.~Talbot}, {\em First-passage times in
  $d$-dimensional heterogeneous media}, Phys Rev Lett, 115 (2015), p.~240601.

\bibitem{viswanathan2008}
{\sc G.~Viswanathan, E.~Raposo, and M.~Da~Luz}, {\em {L}{\'e}vy flights and
  superdiffusion in the context of biological encounters and random searches},
  Physics of Life Reviews, 5 (2008), pp.~133--150.

\bibitem{viswanathan1999}
{\sc G.~M. Viswanathan, S.~V. Buldyrev, S.~Havlin, M.~Da~Luz, E.~Raposo, and
  H.~E. Stanley}, {\em Optimizing the success of random searches}, nature, 401
  (1999), pp.~911--914.

\bibitem{wardak2020}
{\sc A.~Wardak}, {\em First passage leapovers of {L}{\'e}vy flights and the
  proper formulation of absorbing boundary conditions}, Journal of Physics A:
  Mathematical and Theoretical, 53 (2020), p.~375001.

\bibitem{wieschen2020}
{\sc E.~M. Wieschen, A.~Voss, and S.~Radev}, {\em Jumping to conclusion? a
  l{\'e}vy flight model of decision making}, The Quantitative Methods for
  Psychology, 16 (2020), pp.~120--132.

\end{thebibliography}

\appendix

\section{Additional Considerations for the Numerical Discretization of the Periodic Fractional Laplacian}\label{app:num-frac-lap}

To numerically implement \eqref{eq:num-hitting} we choose weights $w_n$ ($n\in\mathbb{Z}$) that are based on linear interpolants. Specifically, we define (see Section 3.1 of \cite{huang_2014})
\begin{equation}
    F(t):= \begin{cases} \frac{C_s}{2s(2s-1)}|t|^{1-2s}, & s\neq 1/2, \\ -C_s\log|t|, & s=1/2,\end{cases}
\end{equation}
where $C_s$ is given by \eqref{eq:frac-lap-def} and in terms of which the weights are given by
\begin{equation}
    w_n := \frac{1}{h^{2s}}\begin{cases} \frac{C_s}{2-2s}-F'(1)+F(2)-F(1), & |n|=1, \\ F(n+1)-2F(n)+F(n-1),& |n|\geq 2.\end{cases}
\end{equation}

We use the explicit form of the weights to numerically speed up the evaluation of the infinite sums appearing in the definition of $W_\sigma$ in \eqref{eq:num-frac-lap}. For sufficiently large $n\in\mathbb{Z}$ we have
\begin{equation*}
    w_n = \frac{C_s}{h^{2s}|n|^{1+2s}}\left(1 + O\left(\frac{1}{n^4}\right) \right),
\end{equation*}
so that for any fixed $\sigma\in\mathbb{Z}$ and any sufficiently large integer $k\geq 1$ we have
\begin{equation*}
    w_{\sigma-kM} + w_{\sigma+kM} = \frac{C_s}{2^{2s-1}Mk^{1+2s}}\left(1 + O\left(\left(\frac{\sigma}{kM}\right)^{2}\right)\right).
\end{equation*}
Choosing a sufficiently large integer $K\geq 1$ we obtain
\begin{equation}\label{eq:weight-fast-summation}
    W_{\sigma} = w_\sigma + \sum_{k=1}^K \left(w_{\sigma-kM}+w_{\sigma+kM}\right) + \frac{C_s}{2^{2s-1} M}\zeta(1+2s,K+1) + O\left(\frac{\sigma^2}{M^3K^{2+2s}}\right),
\end{equation}
where $\zeta(z,q) := \sum_{n=0}^\infty (n+q)^{-z}$ is the Hurwitz zeta function which can be quickly computed by standard numerical libraries.  This formula for the weights $W_\sigma$ $(\sigma\in\mathbb{Z})$ provides a good approximation for $W_\sigma$ for moderately sized $K$ thereby reducing computational costs.

\end{document}